\documentclass[aps,pra,reprint,showpacs,showkeys,amsmath,amssymb,floatfix]{revtex4-2}

\usepackage{graphicx}  
\usepackage{dcolumn}   
\usepackage{bm}        
\usepackage{hyperref}  
\usepackage{physics}   
\usepackage{xcolor}

\begin{document}

\title{Entanglement dynamics of delocalized interacting particles}

\author{M. F. V. Oliveira$^{1}$}
\email{matheus.oliveira@fis.ufal.br}
\author{F. A. B. F. de Moura$^{1}$}
\author{M. L. Lyra$^{1}$}
\author{G. M. A. Almeida$^{1}$}
\affiliation{$^{1}$Instituto de Física, Universidade Federal de Alagoas, 57072-900 Maceió, AL, Brazil}

\date{\today}

\begin{abstract}

Quantum entanglement in systems of identical particles is often obscured by the interplay between exchange-induced correlations and the operational framework used to define entanglement. 
To study the role of exchange statistics, we propose a scheme using two \textit{distinguishable} particles where an exchange symmetry is artificially engineered via a relative phase $\theta$ in the initial state. This approach allows continuous tuning from bosonic ($\theta = 0$) to fermionic ($\theta = \pi$) statistics. By monitoring the interplay between purity and coherence, we uncover distinct dynamical regimes dictated by the interaction strength $U$ and the phase $\theta$. 
For particles initially loaded in a bound state, strong $U$ suppresses coherence development by avoiding the scattering band, reducing the purity toward its minimum. For particles initially on neighboring sites, coherence grows linearly in time. While non-symmetric inputs feature a sharp purity reduction at intermediate $U$, due to the competition between bound and unbound states, symmetric initial conditions produce transient coherence bursts that significantly enhance the purity. More generally, tuning the phase $\theta$ reveals a high-purity region over a range of $\theta$ at intermediate interactions, with the purity collapsing to $1/2$ as $\theta$ approaches the fermionic limit. Our results show that the imposed statistics, or lack thereof, reshapes the entanglement dynamics and its response to the interaction $U$.

\end{abstract}

\maketitle

\section{Introduction}

    Quantum entanglement is a fundamental resource for emerging technologies such as quantum computing and quantum simulation \cite{Nielsen2010,Georgescu2014,Pirandola2015}, and it also serves as an indicator of phase transitions \cite{Osterloh2002,Vidal2003,Gu2004}. This nonclassical property allows particles to share instantaneous correlations; for instance, in a bipartite system the information about entanglement between the subsystems is encoded in the reduced density matrix $\rho_a$. Quantifying this entanglement, however, faces fundamental obstacles when identical and delocalized particles are involved \cite{Islam2015, LoFranco2018}.

    The difficulty arises because indistinguishability introduces a fundamental ambiguity. While entanglement
    between identical particles is proved to be a useful resource \cite{morris20},
    the impossibility of labeling individual particles and their symmetrization requirements blends statistical correlations with 
    operationally accessible entanglement \cite{Killoran2014,LoFranco2016}. The von Neumann entropy, the standard measure for bipartite pure states,
    does not uniquely separate exchange-induced correlations from entanglement defined with respect to distinguishable subsystems \cite{Ghirardi2004,Tichy2011}. Furthermore, it requires the diagonalization of $\rho_a$, which is a costly task even for relatively simple systems. 
    This confusion has led to a proliferation of alternative concepts and measures \cite{Sasaki2011, Balachandran2013, Benatti2014}, many of which remain technically involved and inadequate for quantifying entanglement in realistic scenarios of spatial overlap \cite{Veldhorst2015, LoFranco2016}. Even alternatives proposed to circumvent the computational cost, such as the linear entropy \cite{Zurek1993,Manfredi2000,Morelli2020}, prove problematic in low-density regimes, leading to incorrect predictions about entanglement \cite{Pauletti2024}.

   To study these questions in a controlled setting, we consider a system of two interacting distinguishable particles, where an effective exchange symmetry is artificially controlled by the relative phase $\theta$ in the initial state. 
This key aspect allows us to continuously tune between bosonic ($\theta = 0$) and fermionic ($\theta = \pi$) statistics, isolating the role of exchange symmetry while avoiding the intrinsic indistinguishability of the particles. The phase $\theta$ enables the reproduction of characteristic behaviors such as bunching or antibunching \cite{Lahini2010,Lahini2012,oliveira2023}.
Within this framework, we investigate how correlations develop dynamically under the combined effect of interaction and symmetry. The time evolution plays a central role, as it reveals the mechanisms by which coherence and mixedness are generated, redistributed, and suppressed, including effects associated with bound states and their hybridization with the scattering continuum \cite{Shepelyansky1994, Bromberg2009, Lahini2012, Lee2014, Ferreira2022, oliveira2023}. This model, despite its simplicity, captures a wide range of phenomena and admits direct experimental implementation in photonic lattices \cite{Peruzzo2010, Krimer2011, Longhi2011, Corrielli2013, Lee2014}.

    In this context, the nontrivial association between entanglement and transport phenomena in two-particle systems makes the topic particularly interesting. 
    In order to address the role of the particles’ effective identity in their propagation alongside generation of entanglement we rely on the information available in the single-particle reduced density matrix $\rho_a$.
    %
     Our analysis is made through two complementary quantities: the purity $\gamma_a=\mathrm{Tr}(\rho_a^2)$, which quantifies the statistical mixedness of the reduced state, and the $L_1$-norm of coherence ($C_{L1}$) \cite{Baumgratz2014}, which quantifies the quantum superpositions present in $\rho_a$ in the local (site) basis.
Note that while $\gamma_a$ provides a basis-independent proxy for bipartite entanglement (for global pure states), the $L_1$-norm of coherence is basis dependent. Therefore, these quantities are not intended to decompose entanglement, but rather to provide complementary diagnostics of the underlying dynamical processes in a physically motivated basis.

    With this combination, in the present work we demonstrate that the control of symmetry via $\theta$ (or lack thereof) allows the generation of correlation regimes that are not trivially accessible with identical particles, and it allows us to precisely map the transition between them. 
    In particular, we recover behaviors analogous to those observed in interacting two-particle systems—such as the emergence of interaction-dependent correlations and the role of bound states in shaping the dynamics \cite{Lahini2012}. Moreover, by jointly analyzing purity and coherence, we provide a complementary characterization of the correlations, in the spirit of previous studies where different quantifiers reveal distinct aspects of the underlying quantum state \cite{Ferreira2022}.
    %

    

\section{Methods}

    \subsection{Model and basis decomposition}
    
    Let us start by considering two interacting distinguishable particles moving on a linear chain consisting of $N$ sites. The Hamiltonian of the system is given by $(\hbar = 1)$

        \begin{equation}
            H = J\sum_{i = 1}^{N - 1}\bigl(a_{i+1}^{\dagger}a_{i} + b_{i+1}^{\dagger}b_{i} + \mathrm{h.c.}\bigr)
              + U\sum_{i = 1}^{N}a_{i}^{\dagger}a_{i}\,b_{i}^{\dagger}b_{i}\,,
        \end{equation}
        
    \noindent where $a_i^\dagger$ and $b_i^\dagger$ are the creation operators for each particle at site $i$, $J$ is the nearest-neighbor hopping amplitude, and $U$ is the local (repulsive) interaction strength. 
Here, we adopt $J\equiv 1$ as our standard energy unit.
    We set $J>0$ with no loss of generality. For a bipartite lattice this choise is unitarily equivalent to the standard negative-hopping form, leading only to a mirror symmetry of the energy spectrum, without affecting the eigenstates’ spatial structure or the unitary time evolution up to an overall phase.
    The two-particle Hilbert space is spanned by the $N^2$ basis states $\{\ket{m,n} = b_n^\dagger\,a_m^\dagger\ket{\mathrm{vacuum}}\}_{m,n=1}^{N}$.
    
    A key aspect of our approach is the preparation of the initial state in a general superposition of the form:
    \begin{equation}\label{eq:initial_state}
        \ket{\psi(0)} = \frac{\ket{m_0,n_0} + e^{i\theta}\ket{n_0,m_0}}{\sqrt{2}},
    \end{equation}  
     where the relative phase $\theta$ artificially controls the effective exchange symmetry between the particles. This allows us to continuously tune the initial state from symmetric ($\theta = 0$) to antisymmetric ($\theta = \pi$), mimicking bosonic or fermionic statistics. 

    For the specific cases $\theta = 0$ and $\theta = \pi$, the time-evolved state $\ket{\psi(t)} = e^{-iHt/\hbar}\,\ket{\psi(0)}$ access either one of two decoupled subspaces.
    This is revealed by using the symmetric and antisymmetric basis:
    \begin{equation*}
        \ket{m,n}^{\pm} = \frac{\ket{m,n}\pm\ket{n,m}}{\sqrt{2}}\quad (m\neq n).
    \end{equation*}
    The antisymmetric subspace ($-$) describes noninteracting spinless fermions, while the symmetric subspace ($+$), which also contains the doubly occupied states $\ket{m,m}$, behaves like bosons \cite{Krimer2011,oliveira2023}. Note that the interaction $U$ acts only within the symmetric subspace, since the Pauli exclusion principle forbids double occupancy in the antisymmetric subspace.
This leads us to define the projectors $P_{\pm}$ onto the symmetric and antisymmetric subspaces,
    \begin{equation*}
        P_\pm = \sum_{m<n}\ket{m,n}^\pm\!\bra{m,n}^\pm \;+\;\delta_{\pm,+}\sum_{m}\ket{m,m}\!\bra{m,m},
    \end{equation*}
    \noindent which satisfy $P_+ + P_- = \mathbb{I}$ and $P_+P_- = 0$. This provides a convenient framework to compare how bosonic-like and fermionic-like statistics influence the entanglement dynamics.

\subsection{Measures of coherence and entanglement}

    We now define the quantities used to track entanglement dynamics. As mentioned earlier, we employ two standard measures to quantify the degree of mixture and the presence of quantum superpositions in the reduced state of a single particle (subsystem $a$): the purity and the $L_1$-norm of coherence. These measures are computed from the reduced density matrix: $\rho_a(t) =tr_b[\rho_{ab}(t)]$ where $\rho_{ab}(t) =|\psi(t)\rangle\langle\psi(t)|$ is the full density matrix of the system. 

    The purity is defined as $\gamma_a(t) = \mathrm{Tr}[\rho_a^{2}(t)]$, which quantifies the degree of mixedness of the reduced state. Given the subsystem $a$ has dimension $N$, the purity is bounded by $1/N \le \gamma_a \le 1$. The upper bound, $\gamma_a = 1$, is attained if and only if particle $a$ is in a pure state, indicating no entanglement with particle $b$. Conversely, the minimum value, $\gamma_a = 1/N$, corresponds to the maximally mixed state $\rho_a = \mathbb{I}/N$, which signifies maximum entanglement between the two particles.

    The second measure is the $L_1$-norm of coherence \cite{Baumgratz2014}, 
    which quantifies the weight of the off-diagonal elements in a chosen reference basis. Here, we use the site occupation basis ${|i\rangle}^{N}_{i=1}$, as it is the natural option for analyzing spatial delocalization and coherence. The $L_1$-norm is thus defined as $C_{L1}(t) = \sum_{i \neq j} |\langle i | \rho_a(t) | j \rangle|$ and satisfies the bounds $0 \le C_{L1} \le N - 1$.
    A value of $C_{L1} = 0$ indicates a perfectly diagonal density matrix in the site basis, meaning the state possesses no quantum coherence (an incoherent state). The upper bound is reached for a fully delocalized pure state as $|\psi\rangle = \frac{1}{\sqrt{N}}\sum_{k=1}^{N}|k\rangle$, for which $\rho_a = |\psi\rangle\langle\psi|$ and $|\langle i|\rho_a|j\rangle| = 1/N$ for all $i\neq j$, giving $C_{L1} = \sum_{i\neq j}1/N = N-1$.

    To aid interpretation of the dynamics, it is useful to decompose the purity into contributions from the diagonal and off-diagonal elements of $\rho_a$ in the site basis. Consider a generic reduced state written as $\rho_a = \sum_{i,j=1}^{N} \rho_{ij}\ket{i}\bra{j}.$ Since $\rho_a$ is Hermitian and has unit trace, the purity can be written as $\gamma_a=\mathrm{Tr}(\rho_a^2)=\sum_{i,k}|\rho_{ik}|^2$, or equivalently 
    \begin{equation}\label{rhoa}
        \gamma_a = \underbrace{\sum_{i}|\rho_{ii}|^{2}}_{\gamma_{\mathrm{diag}}}
                 + \underbrace{2\sum_{i\neq k}|\rho_{ik}|^{2}}_{\gamma_{\mathrm{offdiag}}}.
    \end{equation}
    An interesting point in this perspective is that, although $C_{L1}=\sum_{i\neq k}|\rho_{ik}|$ and $\gamma_{\mathrm{offdiag}}$ are distinct quantities, both quantify coherence by adding
    contributions from the off-diagonal elements, the latter weighting these terms quadratically.
%
   We stress that the equation above represents a decomposition of the purity -- not of entanglement -- into diagonal and off-diagonal contributions in the chosen basis. This decomposition serves as a diagnostic tool to distinguish population spreading from coherence redistribution, rather than a basis-independent classification of entanglement.

\subsection{Time–evolution scheme}

    To simulate the dynamics, we numerically solve the time-dependent Schrödinger equation for the two-particle wavefunction. The state vector is represented in the site basis as $|\psi(t)\rangle=\sum_{m,n} f_{m,n}(t)|m,n\rangle$, where $f_{m,n}(t)$ are the time-dependent probability amplitudes. The evolution operator is expanded in a Taylor series to order $l_f$:
    
    \begin{equation}
        \Gamma(\Delta t) = e^{-iH\Delta t/\hbar} \approx 1 + \sum_{l=1}^{l_f} \frac{(-iH\Delta t/\hbar)^l}{l!}.
    \end{equation}

    \noindent In what follows, we set $\hbar = 1$ and measure time in units of $J^{-1}$. The wavefunction at time $t+\Delta t$ is given by $|\psi(t+\Delta t)\rangle = \Gamma(\Delta t)|\psi(t)\rangle$. The method can be used recursively to obtain the wavefunction at time $t$ \cite{Dias2007}. In our simulations, we use $\Delta t = 0.01$ and truncation order $l_f = 20$ which maintains the wavefunction norm conserved to within $1 - \sum_{m,n} |f_{m,n}(t)|^2 \leq 10^{-14}$ throughout the evolution.
    
    In this way, the reduced density-matrix elements are computed directly from the evolved wavefunction coefficients:
    \begin{equation}
        \langle m|\rho_a(t)|m'\rangle = \sum_{n} f_{m,n}(t) f^{*}_{m',n}(t),
    \end{equation}
     from these matrix elements we calculate the purity $\gamma_a(t)$ and the $L_1$-norm of coherence $C_{L1}(t)$. We remark that the time evolution is implemented via a truncated Taylor expansion, avoiding Hamiltonian diagonalization, while purity provides an entanglement proxy also without requiring diagonalization of the reduced density matrix.
    
    With this computational framework established, we proceed to analyze how the interplay between interaction strength $U$ and initial symmetry $\theta$ shapes the purity and coherence dynamics in the following section.


\section{Results}
    
    Before examining the effects of spatial symmetry, we first analyze the system's response under non-symmetric initial conditions, namely $\ket{\psi(0)}=\ket{m_0,n_0}$, and then proceed to controlled symmetric cases to isolate the role of exchange symmetry on transport and entanglement generation.
    
\subsection{Non-symmetric initial condition}

\begin{figure*}[htbp]
        \centering
        \caption{Entanglement dynamics for two distinguishable particles in a lattice with $N = 300$ sites, evolved up to $tJ = 70$. (a–d) Initially bound particles ($m_0 = n_0 = N/2$). (a) Density map of the purity $\gamma_a(t)$ as a function of time $t$ and interaction strength $U/J$. The red region indicates predominantly mixed states, while the small blue area at the beginning of the evolution [as shown in (c)] reveals a regime of high purity. (b) This behavior is further highlighted by plotting the purity as a function of $U$ at early times: for large $U$, the system’s purity approaches 1, whereas for intermediate interaction strengths, the purity rapidly decreases, indicating faster mixing. (d) $L_1$-norm of coherence $C_{L1}(t)$ versus $t$. For $U = 0$, a steady growth is observed, in which case the particles freely propagate and delocalize.
        However, as $U$ increases, it suppresses the coherence toward zero, indicating that the reduced density matrix approaches a diagonal form on the site basis. The purity remains close to 1 in the early stage of the dynamics because of the small effective hopping strength associated to the band of bound states.
        (e–h) Particles loaded in neighboring sites ($m_{0} = N/2$ and $n_{0} = 1 + N/2$). 
        (e,g) The Density map shows that increasing $U$ prevents the state to become strongly mixed over time, except at intermediate $U$, where
        significant purity decrease is observed. (f) Purity versus $U$, averaged over late times ($tJ \to 70$).
        (h) Differently from the bound input, increasing $U$
        saturates the coherence $C_{L1}$ whilee preserving its linear trend.}
        \includegraphics[width=1.\linewidth]{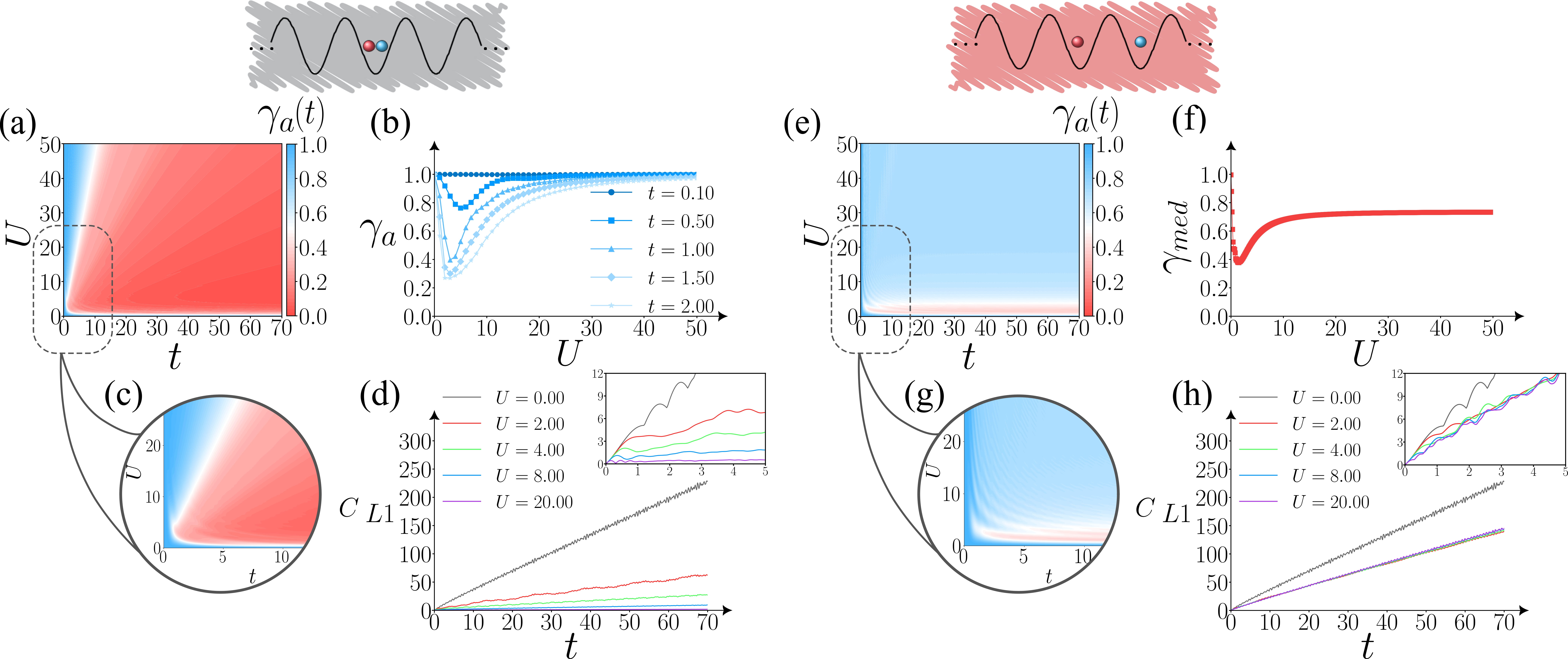}
        \label{figura_prz_dist}
    \end{figure*}

    We begin from two non-symmetric initial configurations: (i) both particles completely localized on the same site ($n_0=m_0$) and (ii) on neighboring sites ($n_0=m_0+1$) as depicted in Figure \ref{figura_prz_dist}. Our goal is to understand how the local interaction $U$ modulates these dynamics.
    For larger initial separations, the reduced overlap between the particles delays the onset of interaction effects, leading to a transient behavior closer to the noninteracting regime before the dynamics qualitatively converge to the behavior discussed below.

    \textit{Particles on the same site.} For non-interacting ($U=0$) bound particles, the system remains in a pure state, reflected in a constant purity $\gamma_a = 1$ [Fig.~\ref{figura_prz_dist}(a)] and a linear growth of coherence $C_{L1}(t)$ [Fig.~\ref{figura_prz_dist}(d)] due to ballistic delocalization \cite{Dias2010}. Introducing interaction ($U > 0$) changes this picture, since 
    the other particle act as an effective environment, resulting in an interaction-induced decoherence process \cite{Zurek1993,Buchleitner2003,Rapedius2012,Benatti2020}. 
    Their interaction also leads to a progressive decay of purity over time. 
    %
    In the strong interaction limit ($U \gg J$), a particularly interesting regime emerges. The development of coherence is suppressed while $\gamma_a(t)$ remains close to 1 for a short time before decaying [Fig.~\ref{figura_prz_dist}(b)]. In this limiting case, 
    the reduced density matrix is diagonal, $\rho_a(t) \sim \sum_{m}|f_{m,m}(t)|^2\ket{m}\bra{m} $, as $U$ forbids access to the scattering band. 
    This behavior is consistent with the strong-coupling picture, where bound states form a narrow band with reduced effective bandwidth, leading to slow dynamics and delayed mixing with unbound states \cite{Lahini2012}.
    
    

    \textit{Particles on neighboring sites.} The dynamics differ qualitatively when particles start on adjacent sites. While $U=0$ again yields constant purity, 
    the increase of interaction leads to a saturation of purity at $\gamma_a \approx 0.7$ that persists in time [see Figs. \ref{figura_prz_dist}(e) and \ref{figura_prz_dist}(f)]. The coherence $C_{L1}$ steadily grows over time but also saturates with $U$ [Fig.~\ref{figura_prz_dist}(h)]. Note that the absence of initial double occupancy prevents the immediate formation of 
    a bound pair.
    This dynamical tradeoff between diagonal and off-diagonal terms of $\gamma_a$ [see Eq. (\ref{rhoa})] rendering a constant purity indicates that both particles are spreading away from each other. Namely, coherence is being generated in the site basis, increasing $\gamma_{\mathrm{offdiag}}$, while $\gamma_\mathrm{diag}$ decreases in a compensating manner. 

The oscillatory behavior observed in the coherence $C_{L1}(t)$ 
originates from the quench nature of the dynamics. 
Since the initial states considered here are not eigenstates of the Hamiltonian, 
they can be written as superpositions of many-body energy eigenstates. During the time evolution, each eigenstate 
component acquires a dynamical phase $e^{-iE_{\alpha}t}$, 
which leads to relative phase differences $e^{-i(E_{\alpha}-E_{\beta})t}$ between distinct contributions. 
These time-dependent phases modulate the off-diagonal elements of the reduced density matrix and therefore generate oscillations in coherence measures such as $C_{L1}(t)$. This mechanism is a general feature of quantum quench dynamics and has been widely discussed in the literature (see, for instance, Ref. \cite{Calabrese2006}).

    In the intermediate $U$ regime, in \emph{both} cases, the hybridization between bound ($n = m$) and unbound ($n \neq m$) states allows significant overlap of their wavefunctions \cite{Dias2010}. This hybridization induces transitions between the two competing state types, 
    redistributing weight between diagonal and off-diagonal contributions, thereby reducing the purity $\gamma_a(t)$ 
    [Figs. \ref{figura_prz_dist}(b) and \ref{figura_prz_dist}(f)]. 
    The main difference is that for bound particles, this purity loss occurs for all interaction values at sufficiently long times [as seen in Fig.~\ref{figura_prz_dist}(a)], 
    while for particles on neighboring sites it is restricted to this intermediate interaction regime [as shown by the thin red region in Fig.~\ref{figura_prz_dist}(e)].



    
    \begin{figure}[h!]
        \centering
        \caption{Phase diagram, parametrized by time (represented by a color gradient), showing the evolution of the purity ($\gamma_a$) and the contribution of the diagonal elements ($\gamma_{\mathrm{diag}}$) of the reduced density matrix for the non-symmetric, bound input, across four different interaction strengths.}
        \includegraphics[width=0.95\linewidth]{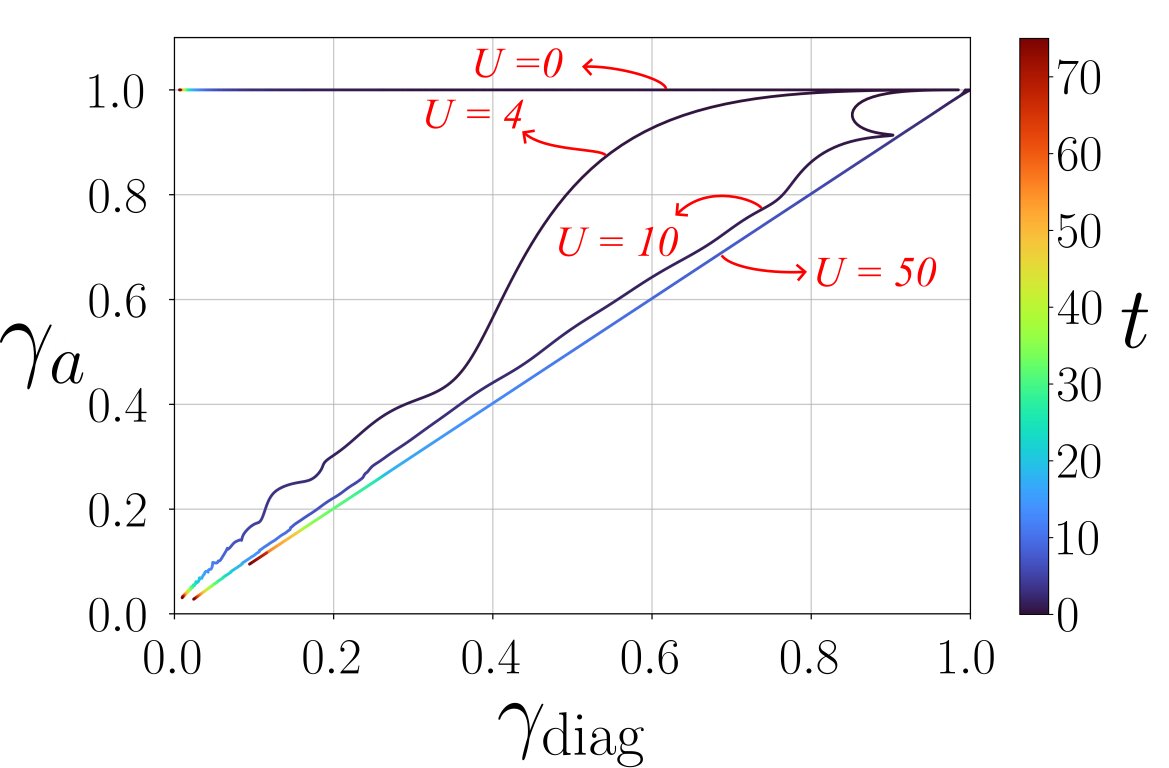}        
        \label{prz_x_przdiag}
    \end{figure}


We can better quantify the interplay between purity and coherence through the phase diagram $\gamma_a \times \gamma_{\text{diag}}$ for the bound input (Fig.~\ref{prz_x_przdiag}). For $U = 0$, the horizontal trajectory ($\gamma_a (t) = 1$; track the color gradient from right to left) reflects perfect balance between diagonal and coherence terms. 
    At $U = 50J$, the linear relation $\gamma_a \approx \gamma_{\text{diag}}$ marks strong suppression of coherence, which plays a negligible role in shaping $\gamma_a$.
    Between these extremes, for intermediate regimes of interaction (e.g., $U = 4J, 10J$ in Fig.~\ref{prz_x_przdiag}), the purity loss 
    is preceded by short bursts of coherence, with the trajectories eventually approaching a linear behavior. 
    Interestingly, for $U = 10J$ the lobe formed early in the dynamics tells us that the system initially gains coherence and shortly after enters a regime 
    where the coherence loss is accompanied by a rapid increase in the diagonal term. 
    

\subsection{Symmetric initial condition}

     %

    \begin{figure}[t!]
        \centering
        \caption{(a,c) Density plot of the time evolution of the purity $\gamma_a(t)$ as a function of $U$ for distinguishable particles with a symmetric initial condition with $\theta=0$ and $n_0=m_0+1$ in Eq. (\ref{eq:initial_state}). (b) $C_{L1}$ as a function of time for selected values of $U$, showing linear growth. 
          During a short transient initial regime and for intermediate values of $U$, we observe an increase in purity [as shown in the zoomed-in panel (c)]. (d) Maximum purity $\gamma_a^{\mathrm{max}}$ obtained during the time evolution as a function of interaction strength $U$.}
        \vspace{0.25cm}
        \includegraphics[width=1.025\linewidth]{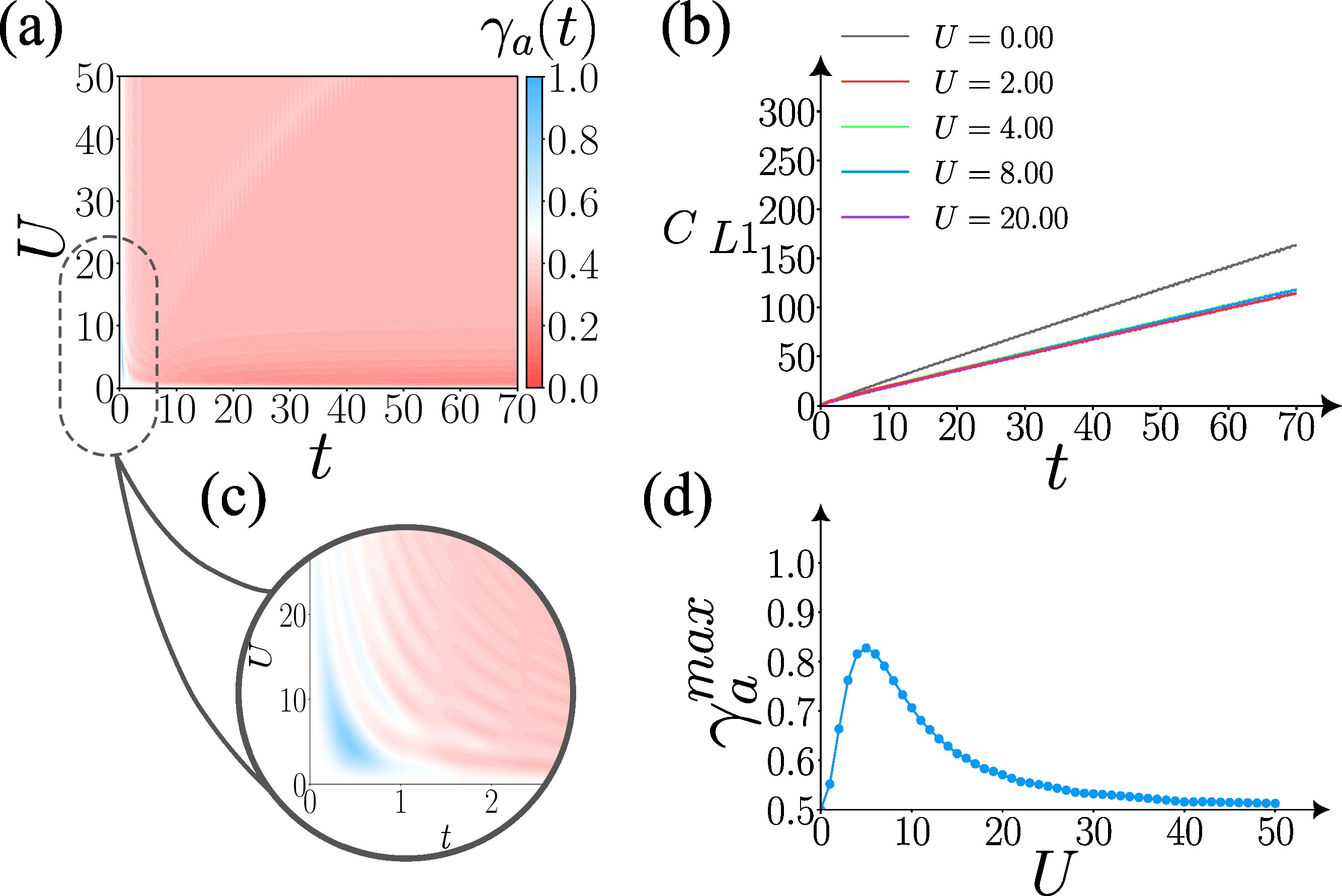}
        \vspace{-0.5cm}
        \label{figura_prz_sym}
    \end{figure}

\begin{figure}[t!]
        \centering
        \caption{Phase diagram, parametrized by time (represented by a color gradient), showing the evolution of the purity ($\gamma_a$) and the contribution of the diagonal elements ($\gamma_{\mathrm{diag}}$) of the reduced density matrix for the symmetric initial condition, as in Fig. \ref{figura_prz_sym}, for four different interaction strengths.}
        \includegraphics[width=0.95\linewidth]{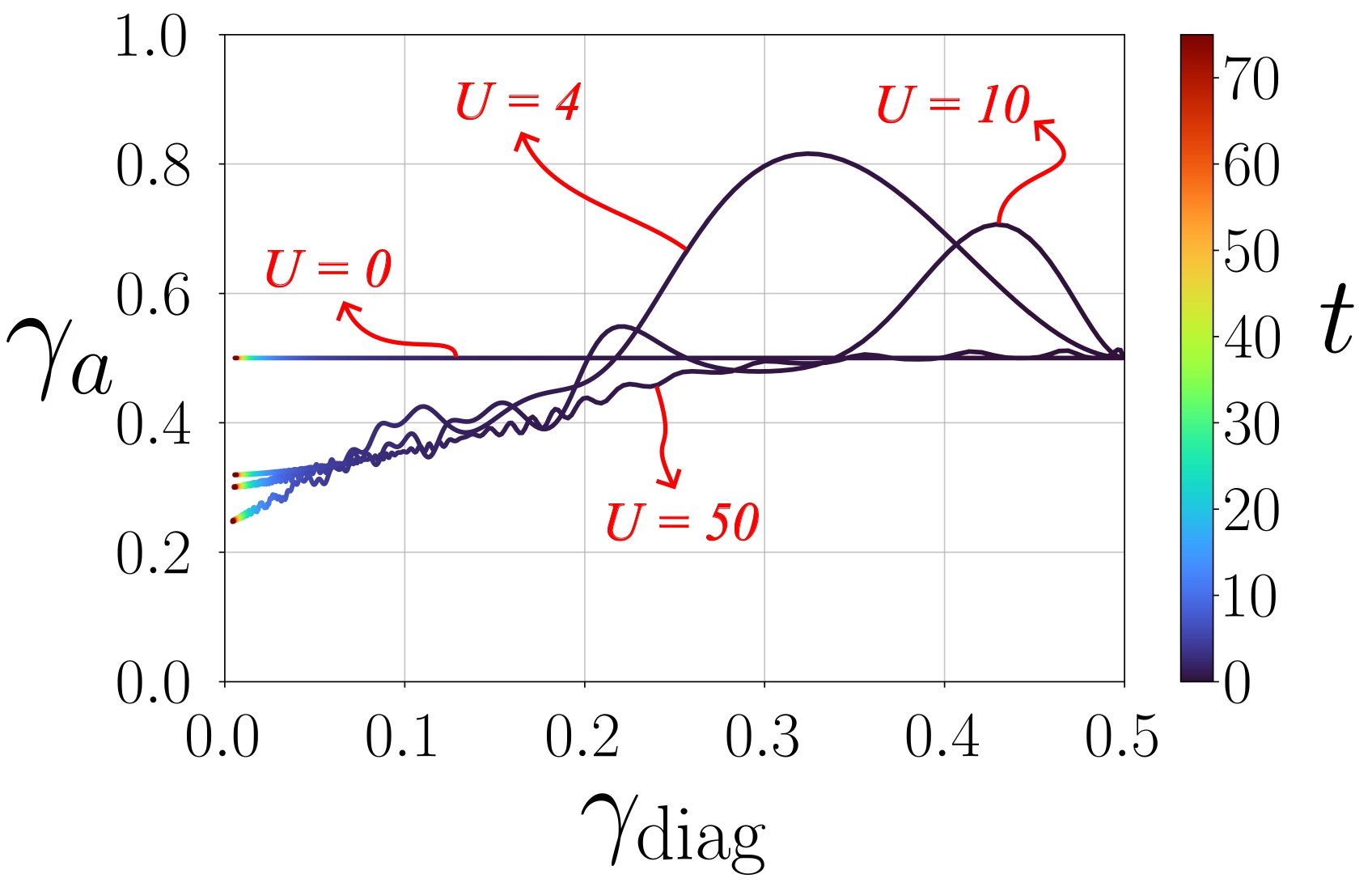}
        \label{prz_x_przdiag_bsl}
    \end{figure}

     We are now ready to examine how the symmetric initial state ($|\psi(t=0)\rangle = \frac{1}{\sqrt{2}}(|m_{0},m_{0}+1\rangle + |m_{0}+1,m_{0}\rangle$) affects the entanglement dynamics of the two particles. We observe in Figure \ref{figura_prz_sym}(a) a
     predominance of mixed states throughout the evolution for practically any interaction strength, with saturation of $\gamma_a$. 
   For $U = 0$, the purity remains constant over time, now with $\gamma_a = 0.5$ (cf. \ref{prz_x_przdiag_bsl}), coinciding with the theoretically expected value for a maximally mixed state in a bipartite system of dimension $d = 2$, where $\gamma_a = 1/d$. 
   Although the Hilbert space is larger ($N > 2$), the system’s dynamics effectively confine the correlations to a lower-dimensional subspace, owing to exchange symmetry, in the absence of interaction.

    The coherence $C_{L1}$ grows linearly for any value of $U$ [Figure~\ref{figura_prz_sym}(b)], analogous to the non-symmetric case [Figure \ref{figura_prz_dist}(h)] due to delocalization of the unbound states. 
    In this scenario, increasing $U$ prevents bound states from participating in the dynamics, what saturates the rate of $C_{L1}$.   
    However, another interesting behavior takes place in the intermediate $U$ regimes, now involving a transient, and significant, increase in the purity at the beginning of the dynamics.
     As the interaction strength increases, purity rises to a maximum at $U \approx 5J$, reaching $\gamma_a^{\max} \approx 0.827$ before decaying again [Figure~\ref{figura_prz_sym}(d)]. 
     This behavior results from the dynamic competition between the hopping and interaction terms. This competition generates constructive interference in the vicinity of $m_0$ due to the hybridization 
     of bound and unbound modes. But then again, the delocalization diminishes the contribution of $\gamma_{\mathrm{diag}}$, ultimately bringing $\gamma_a$ to a saturation point.  

    In the phase diagram of $\gamma_a \times \gamma_{\text{diag}}$ in Figure~\ref{prz_x_przdiag_bsl}, we can observe this behavior in more detail. Despite the input being in a mixed state ($\gamma_a=0.5$), for intermediate interaction regimes (e.g., $U = 4$ and $U = 10$), the dynamics exhibit transient gains of coherence and sudden increase of $\gamma_a$, whereas $\gamma_{\mathrm{diag}}$ is always decreasing.
    For strong interactions ($U = 50$), the trajectory temporarily approximates the behavior of $U = 0$ at short times. In this regime, the interaction prevents the formation of bound states. As $C_{L1}$ steadily increases at a slower rate [Fig. \ref{figura_prz_dist}(h)], a smaller $\gamma_a<0.5$ is achieved as time progresses.

    Comparing the symmetric and non-symmetric cases of both particles loaded in neighboring sites, the bosonic character of the input changes the characteristic purity observed during the dynamics. Most importantly, it dramatically modifies the response to the interaction $U$, as discussed above. Although both particles are distinguishable, the imposed statistics has direct consequences on the entanglement shared between them, as captured by the single-particle reduced density matrix in the position basis and by the steady growth of coherence \cite{Streltsov2015}.

\subsection{General initial condition}

    To generalize our findings, we now consider the initial state given by Eq.~\ref{eq:initial_state}, where the relative phase $\theta$ allows for continuous interpolation between effective bosonic ($\theta = 0$) and fermionic ($\theta = \pi$) statistics. This approach enables us to map how the artificial exchange symmetry influences the maximum attainable purity throughout the dynamics.

\begin{figure}[t!]
        \centering
        \caption{Maximum purity $\gamma_a^{max}$ reached during the time evolution, up to $tJ=70$, as a function of the interaction strength and the input phase $\theta$ within the interval $[0,\pi]$.}
        \vspace{-0.10cm}
        \includegraphics[width=0.9\linewidth]{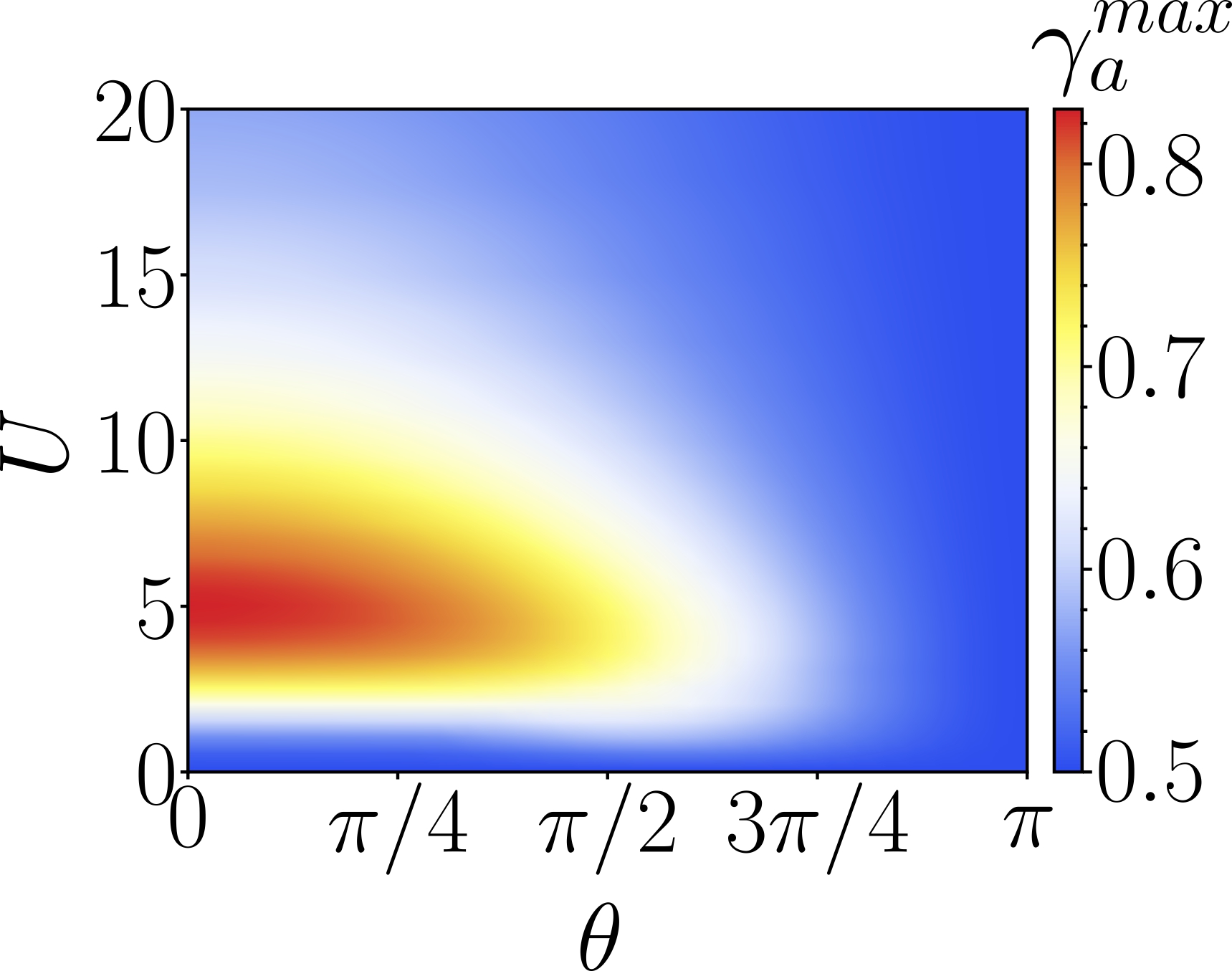}
        \vspace{-0.5cm}
        \label{mapa_przmax}
    \end{figure}

    In Figure~\ref{mapa_przmax}, we present the maximum purity, $\gamma_a^{\mathrm{max}}$, achieved during the time evolution as a function of the initial phase $\theta$ and the interaction strength $U$. The most prominent feature is a well-defined region of high purity for ($0\le \theta \le \pi/2$) and intermediate interactions ($2\lesssim U/J \lesssim 10$).

    This peak can be understood as a signature of synthetic bunching. The constructive interference between the $|m_{0},n_{0}\rangle$ and $|n_{0},m_{0}\rangle$ configurations enhances the probability amplitude for the particles to be found together, reinforcing local coherence and leading to a transient state of higher purity. This effect synergizes with the interaction $U$, which, at intermediate strengths, allow hybridization between bound and unbound states.

    As $\theta$ approaches $\pi$, the interference becomes destructive, mimicking fermionic antibunching. This leads to a monotonic decrease in $\gamma_a^{\mathrm{max}}$. In the fully antisymmetric case ($\theta = \pi$), the maximum purity remains at 0.5, indicating that the dynamics is effectively confined to the antisymmetric subspace that does not suffer influence from $U$. This is consistent with the Pauli exclusion principle, which forbids double-occupied fermionic states.

    Further analysis of Fig. \ref{mapa_przmax} reveals three distinct dynamical regimes. For weak interactions $(U \lesssim 2)$, the particles propagate almost independently and display the similar behavior to the symmetric case ($\theta = 0$), for any $\theta$, since the interaction is too weak to significantly alter the initial correlations. 
    The intermediate interaction regime ($2\lesssim U/J \lesssim 10$) is where the phase $\theta$ plays its most decisive role, as explained above, as the competition between hopping and interaction is most effective.
    Finally, for strong interactions, the band of bound states splits from the scattering band, which then governs the propagation. This slows down the rate of coherence growth, locking the system into an inherently mixed state regardless of the initial phase $\theta$.

\section{Conclusion}

    In summary, we found that purity and coherence are distinctly modulated by the artificially imposed initial symmetry, which qualitatively determines the entanglement dynamics. 
    In the non-symmetric case, when the particles start in a bound state, we observe the suppression of coherence development for $U \gg J$ as the scattering band is avoided during the evolution. It also results in a slower 
    bound-pair
propagation due to the decrease of the effective hopping strength of the associated band. In this regime the purity $\gamma_a$ is dominated by its diagonal part $\gamma_\mathrm{diag}$ such that delocalization brings $\gamma_a$ near its minimum. 
For the particles initially loaded in neighboring sites, the coherence always increases linearly with time, for any value of $U$. 
Although $\gamma_\mathrm{diag}$ inevitably decreases at long times to particle spreading, the overall purity $\gamma_a$ remains comparatively higher. 

A common trait found for both non-symmetric cases is a sharp purity decrease for intermediate values of $U$ due to the hybridization between bound and unbound states. In contrast, for both particles prepared in a symmetric bosonic state, we observe transient bursts of coherence that increase the overall purity. Despite the distinguishability of the particles, the imposed statistics reshapes the entanglement dynamics, the attainable values of purity, and the response to the interaction $U$.
    
    The general initial condition in which the relative phase $\theta$ continuously interpolates between bosonic statistics ($\theta=0$, which recovers the symmetric case) and fermionic statistics ($\theta=\pi$) produces, respectively, particle bunching or antibunching. Mapping the maximum purity as a function of $\theta$ and the interaction strength $U$, we observe a region high purity for $0 \le \theta \lesssim \pi/2$ in the intermediate-interaction regime, peaked around $\gamma_a^{\max}\approx 0.85$. As $\theta$ approaches $\pi$, destructive interference reduces the maximum purity to values close to $1/2$, as established by the input. 

    These results could be tested in experimental platforms such as photonic lattices (e.g., a square array of waveguides) using only classical light sources \cite{Peruzzo2010,Longhi2011,Corrielli2013}; 
    capacitively or inductively coupled superconducting qubits, where the two-particle dynamics are mapped to controllable excitation states \cite{Salathe2015,yan19}, and
    ultracold atoms in 1D optical lattices, where Hubbard interactions can be tuned via Feshbach resonances \cite{Bloch2005,Gross2017}.
In these settings, in particular, repeated projective measurements yield site-resolved occupation snapshots, from which one directly obtains the diagonal elements of the density matrix and their fluctuations, allowing the evaluation of $\gamma_{\mathrm{diag}}$. The off-diagonal elements, which determine both $\gamma_a$ and $C_{L1}$, are encoded in spatial correlations and can be accessed, in principle, via interferometric schemes that map phase coherence onto measurable density correlations.

    As next steps, we propose extending the study to systems with more particles, where the richer structure of permutation symmetry may reveal new propagation patterns and also introducing disorder in the couplings or the on-site potential to verify the robustness of symmetry-induced correlations.
Another interesting perspective for future work concerns the use of overlap-based diagnostics, such as the fidelity or the Loschmidt echo. These quantities measure the overlap between the time-evolved state and the initial state and are widely used to characterize nonequilibrium dynamics and quantum quenches. In the present two-particle system, such measures could provide complementary insight into the dynamical regimes identified here, for example, by revealing how the interplay between interaction strength and exchange phase affects dynamical revivals and the stability of the initial configuration. In this sense, the present minimal two-particle setting may help clarify mechanisms that are also relevant in more complex nonequilibrium many-body dynamics, where Loschmidt echo diagnostics are commonly employed \cite{Quan2010}.

\section*{Acknowledgements}
We thank the financial support from CAPES (Coordenação de Aperfei\c{c}oamento de Pessoal de Nível Superior), CNPq (Conselho Nacional de Desenvolvimento Cient\'ifico e Tecnológico), and FAPEAL (Funda\c{c}\~ao de Apoio \`a  Pesquisa do Estado de Alagoas).


\begin{thebibliography}{45}%
\makeatletter
\providecommand \@ifxundefined [1]{%
 \@ifx{#1\undefined}
}%
\providecommand \@ifnum [1]{%
 \ifnum #1\expandafter \@firstoftwo
 \else \expandafter \@secondoftwo
 \fi
}%
\providecommand \@ifx [1]{%
 \ifx #1\expandafter \@firstoftwo
 \else \expandafter \@secondoftwo
 \fi
}%
\providecommand \natexlab [1]{#1}%
\providecommand \enquote  [1]{``#1''}%
\providecommand \bibnamefont  [1]{#1}%
\providecommand \bibfnamefont [1]{#1}%
\providecommand \citenamefont [1]{#1}%
\providecommand \href@noop [0]{\@secondoftwo}%
\providecommand \href [0]{\begingroup \@sanitize@url \@href}%
\providecommand \@href[1]{\@@startlink{#1}\@@href}%
\providecommand \@@href[1]{\endgroup#1\@@endlink}%
\providecommand \@sanitize@url [0]{\catcode `\\12\catcode `\$12\catcode `\&12\catcode `\#12\catcode `\^12\catcode `\_12\catcode `\%12\relax}%
\providecommand \@@startlink[1]{}%
\providecommand \@@endlink[0]{}%
\providecommand \url  [0]{\begingroup\@sanitize@url \@url }%
\providecommand \@url [1]{\endgroup\@href {#1}{\urlprefix }}%
\providecommand \urlprefix  [0]{URL }%
\providecommand \Eprint [0]{\href }%
\providecommand \doibase [0]{https://doi.org/}%
\providecommand \selectlanguage [0]{\@gobble}%
\providecommand \bibinfo  [0]{\@secondoftwo}%
\providecommand \bibfield  [0]{\@secondoftwo}%
\providecommand \translation [1]{[#1]}%
\providecommand \BibitemOpen [0]{}%
\providecommand \bibitemStop [0]{}%
\providecommand \bibitemNoStop [0]{.\EOS\space}%
\providecommand \EOS [0]{\spacefactor3000\relax}%
\providecommand \BibitemShut  [1]{\csname bibitem#1\endcsname}%
\let\auto@bib@innerbib\@empty
\bibitem [{\citenamefont {Nielsen}\ and\ \citenamefont {Chuang}(2010)}]{Nielsen2010}%
  \BibitemOpen
  \bibfield  {author} {\bibinfo {author} {\bibfnamefont {M.~A.}\ \bibnamefont {Nielsen}}\ and\ \bibinfo {author} {\bibfnamefont {I.~L.}\ \bibnamefont {Chuang}},\ }\href@noop {} {\emph {\bibinfo {title} {Quantum Computation and Quantum Information: 10th Anniversary Edition}}}\ (\bibinfo  {publisher} {Cambridge University Press},\ \bibinfo {year} {2010})\BibitemShut {NoStop}%
\bibitem [{\citenamefont {Georgescu}\ \emph {et~al.}(2014)\citenamefont {Georgescu}, \citenamefont {Ashhab},\ and\ \citenamefont {Nori}}]{Georgescu2014}%
  \BibitemOpen
  \bibfield  {author} {\bibinfo {author} {\bibfnamefont {I.~M.}\ \bibnamefont {Georgescu}}, \bibinfo {author} {\bibfnamefont {S.}~\bibnamefont {Ashhab}},\ and\ \bibinfo {author} {\bibfnamefont {F.}~\bibnamefont {Nori}},\ }\bibfield  {title} {\bibinfo {title} {Quantum simulation},\ }\href {https://doi.org/10.1103/RevModPhys.86.153} {\bibfield  {journal} {\bibinfo  {journal} {Rev. Mod. Phys.}\ }\textbf {\bibinfo {volume} {86}},\ \bibinfo {pages} {153} (\bibinfo {year} {2014})}\BibitemShut {NoStop}%
\bibitem [{\citenamefont {Pirandola}\ \emph {et~al.}(2015)\citenamefont {Pirandola}, \citenamefont {Eisert}, \citenamefont {Weedbrook}, \citenamefont {Furusawa},\ and\ \citenamefont {Braunstein}}]{Pirandola2015}%
  \BibitemOpen
  \bibfield  {author} {\bibinfo {author} {\bibfnamefont {S.}~\bibnamefont {Pirandola}}, \bibinfo {author} {\bibfnamefont {J.}~\bibnamefont {Eisert}}, \bibinfo {author} {\bibfnamefont {C.}~\bibnamefont {Weedbrook}}, \bibinfo {author} {\bibfnamefont {A.}~\bibnamefont {Furusawa}},\ and\ \bibinfo {author} {\bibfnamefont {S.~L.}\ \bibnamefont {Braunstein}},\ }\bibfield  {title} {\bibinfo {title} {Advances in quantum teleportation},\ }\href {https://doi.org/10.1038/nphoton.2015.154} {\bibfield  {journal} {\bibinfo  {journal} {Nature Photonics}\ }\textbf {\bibinfo {volume} {9}},\ \bibinfo {pages} {641} (\bibinfo {year} {2015})}\BibitemShut {NoStop}%
\bibitem [{\citenamefont {Osterloh}\ \emph {et~al.}(2002)\citenamefont {Osterloh}, \citenamefont {Amico}, \citenamefont {Falci},\ and\ \citenamefont {Fazio}}]{Osterloh2002}%
  \BibitemOpen
  \bibfield  {author} {\bibinfo {author} {\bibfnamefont {A.}~\bibnamefont {Osterloh}}, \bibinfo {author} {\bibfnamefont {L.}~\bibnamefont {Amico}}, \bibinfo {author} {\bibfnamefont {G.}~\bibnamefont {Falci}},\ and\ \bibinfo {author} {\bibfnamefont {R.}~\bibnamefont {Fazio}},\ }\bibfield  {title} {\bibinfo {title} {Scaling of entanglement close to a quantum phase transition},\ }\href {https://doi.org/10.1038/416608a} {\bibfield  {journal} {\bibinfo  {journal} {Nature}\ }\textbf {\bibinfo {volume} {416}},\ \bibinfo {pages} {608} (\bibinfo {year} {2002})}\BibitemShut {NoStop}%
\bibitem [{\citenamefont {Vidal}\ \emph {et~al.}(2003)\citenamefont {Vidal}, \citenamefont {Latorre}, \citenamefont {Rico},\ and\ \citenamefont {Kitaev}}]{Vidal2003}%
  \BibitemOpen
  \bibfield  {author} {\bibinfo {author} {\bibfnamefont {G.}~\bibnamefont {Vidal}}, \bibinfo {author} {\bibfnamefont {J.~I.}\ \bibnamefont {Latorre}}, \bibinfo {author} {\bibfnamefont {E.}~\bibnamefont {Rico}},\ and\ \bibinfo {author} {\bibfnamefont {A.}~\bibnamefont {Kitaev}},\ }\bibfield  {title} {\bibinfo {title} {Entanglement in quantum critical phenomena},\ }\href {https://doi.org/10.1103/PhysRevLett.90.227902} {\bibfield  {journal} {\bibinfo  {journal} {Phys. Rev. Lett.}\ }\textbf {\bibinfo {volume} {90}},\ \bibinfo {pages} {227902} (\bibinfo {year} {2003})}\BibitemShut {NoStop}%
\bibitem [{\citenamefont {Gu}\ \emph {et~al.}(2004)\citenamefont {Gu}, \citenamefont {Deng}, \citenamefont {Li},\ and\ \citenamefont {Lin}}]{Gu2004}%
  \BibitemOpen
  \bibfield  {author} {\bibinfo {author} {\bibfnamefont {S.-J.}\ \bibnamefont {Gu}}, \bibinfo {author} {\bibfnamefont {S.-S.}\ \bibnamefont {Deng}}, \bibinfo {author} {\bibfnamefont {Y.-Q.}\ \bibnamefont {Li}},\ and\ \bibinfo {author} {\bibfnamefont {H.-Q.}\ \bibnamefont {Lin}},\ }\bibfield  {title} {\bibinfo {title} {Entanglement and quantum phase transition in the extended hubbard model},\ }\href {https://doi.org/10.1103/PhysRevLett.93.086402} {\bibfield  {journal} {\bibinfo  {journal} {Phys. Rev. Lett.}\ }\textbf {\bibinfo {volume} {93}},\ \bibinfo {pages} {086402} (\bibinfo {year} {2004})}\BibitemShut {NoStop}%
\bibitem [{\citenamefont {Islam}\ \emph {et~al.}(2015)\citenamefont {Islam}, \citenamefont {Ma}, \citenamefont {Preiss}, \citenamefont {Tai}, \citenamefont {Lukin}, \citenamefont {Rispoli},\ and\ \citenamefont {Greiner}}]{Islam2015}%
  \BibitemOpen
  \bibfield  {author} {\bibinfo {author} {\bibfnamefont {R.}~\bibnamefont {Islam}}, \bibinfo {author} {\bibfnamefont {R.}~\bibnamefont {Ma}}, \bibinfo {author} {\bibfnamefont {P.~M.}\ \bibnamefont {Preiss}}, \bibinfo {author} {\bibfnamefont {M.~E.}\ \bibnamefont {Tai}}, \bibinfo {author} {\bibfnamefont {A.}~\bibnamefont {Lukin}}, \bibinfo {author} {\bibfnamefont {M.}~\bibnamefont {Rispoli}},\ and\ \bibinfo {author} {\bibfnamefont {M.}~\bibnamefont {Greiner}},\ }\bibfield  {title} {\bibinfo {title} {Measuring entanglement entropy in a quantum many-body system},\ }\href {https://doi.org/10.1038/nature15750} {\bibfield  {journal} {\bibinfo  {journal} {Nature}\ }\textbf {\bibinfo {volume} {528}},\ \bibinfo {pages} {77} (\bibinfo {year} {2015})}\BibitemShut {NoStop}%
\bibitem [{\citenamefont {Lo~Franco}\ and\ \citenamefont {Compagno}(2018)}]{LoFranco2018}%
  \BibitemOpen
  \bibfield  {author} {\bibinfo {author} {\bibfnamefont {R.}~\bibnamefont {Lo~Franco}}\ and\ \bibinfo {author} {\bibfnamefont {G.}~\bibnamefont {Compagno}},\ }\bibfield  {title} {\bibinfo {title} {Indistinguishability of elementary systems as a resource for quantum information processing},\ }\href {https://doi.org/10.1103/PhysRevLett.120.240403} {\bibfield  {journal} {\bibinfo  {journal} {Phys. Rev. Lett.}\ }\textbf {\bibinfo {volume} {120}},\ \bibinfo {pages} {240403} (\bibinfo {year} {2018})}\BibitemShut {NoStop}%
\bibitem [{\citenamefont {Morris}\ \emph {et~al.}(2020)\citenamefont {Morris}, \citenamefont {Yadin}, \citenamefont {Fadel}, \citenamefont {Zibold}, \citenamefont {Treutlein},\ and\ \citenamefont {Adesso}}]{morris20}%
  \BibitemOpen
  \bibfield  {author} {\bibinfo {author} {\bibfnamefont {B.}~\bibnamefont {Morris}}, \bibinfo {author} {\bibfnamefont {B.}~\bibnamefont {Yadin}}, \bibinfo {author} {\bibfnamefont {M.}~\bibnamefont {Fadel}}, \bibinfo {author} {\bibfnamefont {T.}~\bibnamefont {Zibold}}, \bibinfo {author} {\bibfnamefont {P.}~\bibnamefont {Treutlein}},\ and\ \bibinfo {author} {\bibfnamefont {G.}~\bibnamefont {Adesso}},\ }\bibfield  {title} {\bibinfo {title} {Entanglement between identical particles is a useful and consistent resource},\ }\href {https://doi.org/10.1103/PhysRevX.10.041012} {\bibfield  {journal} {\bibinfo  {journal} {Phys. Rev. X}\ }\textbf {\bibinfo {volume} {10}},\ \bibinfo {pages} {041012} (\bibinfo {year} {2020})}\BibitemShut {NoStop}%
\bibitem [{\citenamefont {Killoran}\ \emph {et~al.}(2014)\citenamefont {Killoran}, \citenamefont {Cramer},\ and\ \citenamefont {Plenio}}]{Killoran2014}%
  \BibitemOpen
  \bibfield  {author} {\bibinfo {author} {\bibfnamefont {N.}~\bibnamefont {Killoran}}, \bibinfo {author} {\bibfnamefont {M.}~\bibnamefont {Cramer}},\ and\ \bibinfo {author} {\bibfnamefont {M.~B.}\ \bibnamefont {Plenio}},\ }\bibfield  {title} {\bibinfo {title} {Extracting entanglement from identical particles},\ }\href {https://doi.org/10.1103/PhysRevLett.112.150501} {\bibfield  {journal} {\bibinfo  {journal} {Physical Review Letters}\ }\textbf {\bibinfo {volume} {112}},\ \bibinfo {pages} {150501} (\bibinfo {year} {2014})}\BibitemShut {NoStop}%
\bibitem [{\citenamefont {Lo~Franco}\ and\ \citenamefont {Compagno}(2016)}]{LoFranco2016}%
  \BibitemOpen
  \bibfield  {author} {\bibinfo {author} {\bibfnamefont {R.}~\bibnamefont {Lo~Franco}}\ and\ \bibinfo {author} {\bibfnamefont {G.}~\bibnamefont {Compagno}},\ }\bibfield  {title} {\bibinfo {title} {Quantum entanglement of identical particles by standard information-theoretic notions},\ }\href {https://doi.org/10.1038/srep20603} {\bibfield  {journal} {\bibinfo  {journal} {Scientific Reports}\ }\textbf {\bibinfo {volume} {6}},\ \bibinfo {pages} {20603} (\bibinfo {year} {2016})}\BibitemShut {NoStop}%
\bibitem [{\citenamefont {Ghirardi}\ and\ \citenamefont {Marinatto}(2004)}]{Ghirardi2004}%
  \BibitemOpen
  \bibfield  {author} {\bibinfo {author} {\bibfnamefont {G.}~\bibnamefont {Ghirardi}}\ and\ \bibinfo {author} {\bibfnamefont {L.}~\bibnamefont {Marinatto}},\ }\bibfield  {title} {\bibinfo {title} {General criterion for the entanglement of two indistinguishable particles},\ }\href {https://doi.org/10.1103/PhysRevA.70.012109} {\bibfield  {journal} {\bibinfo  {journal} {Phys. Rev. A}\ }\textbf {\bibinfo {volume} {70}},\ \bibinfo {pages} {012109} (\bibinfo {year} {2004})}\BibitemShut {NoStop}%
\bibitem [{\citenamefont {Tichy}\ \emph {et~al.}(2011)\citenamefont {Tichy}, \citenamefont {Mintert},\ and\ \citenamefont {Buchleitner}}]{Tichy2011}%
  \BibitemOpen
  \bibfield  {author} {\bibinfo {author} {\bibfnamefont {M.~C.}\ \bibnamefont {Tichy}}, \bibinfo {author} {\bibfnamefont {F.}~\bibnamefont {Mintert}},\ and\ \bibinfo {author} {\bibfnamefont {A.}~\bibnamefont {Buchleitner}},\ }\bibfield  {title} {\bibinfo {title} {Essential entanglement for atomic and molecular physics},\ }\href {https://doi.org/10.1088/0953-4075/44/19/192001} {\bibfield  {journal} {\bibinfo  {journal} {Journal of Physics B: Atomic, Molecular and Optical Physics}\ }\textbf {\bibinfo {volume} {44}},\ \bibinfo {pages} {192001} (\bibinfo {year} {2011})}\BibitemShut {NoStop}%
\bibitem [{\citenamefont {Sasaki}\ \emph {et~al.}(2011)\citenamefont {Sasaki}, \citenamefont {Ichikawa},\ and\ \citenamefont {Tsutsui}}]{Sasaki2011}%
  \BibitemOpen
  \bibfield  {author} {\bibinfo {author} {\bibfnamefont {T.}~\bibnamefont {Sasaki}}, \bibinfo {author} {\bibfnamefont {T.}~\bibnamefont {Ichikawa}},\ and\ \bibinfo {author} {\bibfnamefont {I.}~\bibnamefont {Tsutsui}},\ }\bibfield  {title} {\bibinfo {title} {Entanglement of indistinguishable particles},\ }\href {https://doi.org/10.1103/PhysRevA.83.012113} {\bibfield  {journal} {\bibinfo  {journal} {Physical Review A}\ }\textbf {\bibinfo {volume} {83}},\ \bibinfo {pages} {012113} (\bibinfo {year} {2011})}\BibitemShut {NoStop}%
\bibitem [{\citenamefont {Balachandran}\ \emph {et~al.}(2013)\citenamefont {Balachandran}, \citenamefont {Govindarajan}, \citenamefont {de~Queiroz},\ and\ \citenamefont {Reyes-Lega}}]{Balachandran2013}%
  \BibitemOpen
  \bibfield  {author} {\bibinfo {author} {\bibfnamefont {A.~P.}\ \bibnamefont {Balachandran}}, \bibinfo {author} {\bibfnamefont {T.~R.}\ \bibnamefont {Govindarajan}}, \bibinfo {author} {\bibfnamefont {A.~R.}\ \bibnamefont {de~Queiroz}},\ and\ \bibinfo {author} {\bibfnamefont {A.~F.}\ \bibnamefont {Reyes-Lega}},\ }\bibfield  {title} {\bibinfo {title} {Entanglement and particle identity: a unifying approach},\ }\href {https://doi.org/10.1103/PhysRevLett.110.080503} {\bibfield  {journal} {\bibinfo  {journal} {Physical Review Letters}\ }\textbf {\bibinfo {volume} {110}},\ \bibinfo {pages} {080503} (\bibinfo {year} {2013})}\BibitemShut {NoStop}%
\bibitem [{\citenamefont {Benatti}\ \emph {et~al.}(2014)\citenamefont {Benatti}, \citenamefont {Floreanini},\ and\ \citenamefont {Titimbo}}]{Benatti2014}%
  \BibitemOpen
  \bibfield  {author} {\bibinfo {author} {\bibfnamefont {F.}~\bibnamefont {Benatti}}, \bibinfo {author} {\bibfnamefont {R.}~\bibnamefont {Floreanini}},\ and\ \bibinfo {author} {\bibfnamefont {K.}~\bibnamefont {Titimbo}},\ }\bibfield  {title} {\bibinfo {title} {Entanglement of identical particles},\ }\href {https://doi.org/10.1142/S1230161214400031} {\bibfield  {journal} {\bibinfo  {journal} {Open Systems and Information Dynamics}\ }\textbf {\bibinfo {volume} {21}},\ \bibinfo {pages} {1440003} (\bibinfo {year} {2014})}\BibitemShut {NoStop}%
\bibitem [{\citenamefont {Veldhorst}\ \emph {et~al.}(2015)\citenamefont {Veldhorst}, \citenamefont {Yang}, \citenamefont {Hwang}, \citenamefont {Huang}, \citenamefont {Dehollain}, \citenamefont {Muhonen}, \citenamefont {Simmons}, \citenamefont {Laucht}, \citenamefont {Hudson}, \citenamefont {Itoh}, \citenamefont {Morello},\ and\ \citenamefont {Dzurak}}]{Veldhorst2015}%
  \BibitemOpen
  \bibfield  {author} {\bibinfo {author} {\bibfnamefont {M.}~\bibnamefont {Veldhorst}}, \bibinfo {author} {\bibfnamefont {C.~H.}\ \bibnamefont {Yang}}, \bibinfo {author} {\bibfnamefont {J.~C.~C.}\ \bibnamefont {Hwang}}, \bibinfo {author} {\bibfnamefont {W.}~\bibnamefont {Huang}}, \bibinfo {author} {\bibfnamefont {J.~P.}\ \bibnamefont {Dehollain}}, \bibinfo {author} {\bibfnamefont {J.~T.}\ \bibnamefont {Muhonen}}, \bibinfo {author} {\bibfnamefont {S.}~\bibnamefont {Simmons}}, \bibinfo {author} {\bibfnamefont {A.}~\bibnamefont {Laucht}}, \bibinfo {author} {\bibfnamefont {F.~E.}\ \bibnamefont {Hudson}}, \bibinfo {author} {\bibfnamefont {K.~M.}\ \bibnamefont {Itoh}}, \bibinfo {author} {\bibfnamefont {A.}~\bibnamefont {Morello}},\ and\ \bibinfo {author} {\bibfnamefont {A.~S.}\ \bibnamefont {Dzurak}},\ }\bibfield  {title} {\bibinfo {title} {A two-qubit logic gate in silicon},\ }\href {https://doi.org/10.1038/nature15263} {\bibfield  {journal} {\bibinfo  {journal} {Nature}\ }\textbf {\bibinfo {volume} {526}},\
  \bibinfo {pages} {410} (\bibinfo {year} {2015})}\BibitemShut {NoStop}%
\bibitem [{\citenamefont {Zurek}\ \emph {et~al.}(1993)\citenamefont {Zurek}, \citenamefont {Habib},\ and\ \citenamefont {Paz}}]{Zurek1993}%
  \BibitemOpen
  \bibfield  {author} {\bibinfo {author} {\bibfnamefont {W.~H.}\ \bibnamefont {Zurek}}, \bibinfo {author} {\bibfnamefont {S.}~\bibnamefont {Habib}},\ and\ \bibinfo {author} {\bibfnamefont {J.~P.}\ \bibnamefont {Paz}},\ }\bibfield  {title} {\bibinfo {title} {Coherent states via decoherence},\ }\href@noop {} {\bibfield  {journal} {\bibinfo  {journal} {Physical Review Letters}\ }\textbf {\bibinfo {volume} {70}},\ \bibinfo {pages} {1187} (\bibinfo {year} {1993})}\BibitemShut {NoStop}%
\bibitem [{\citenamefont {Manfredi}\ and\ \citenamefont {Feix}(2000)}]{Manfredi2000}%
  \BibitemOpen
  \bibfield  {author} {\bibinfo {author} {\bibfnamefont {G.}~\bibnamefont {Manfredi}}\ and\ \bibinfo {author} {\bibfnamefont {M.~R.}\ \bibnamefont {Feix}},\ }\bibfield  {title} {\bibinfo {title} {Entropy and wigner functions},\ }\href@noop {} {\bibfield  {journal} {\bibinfo  {journal} {Physical Review E}\ }\textbf {\bibinfo {volume} {62}},\ \bibinfo {pages} {4665} (\bibinfo {year} {2000})}\BibitemShut {NoStop}%
\bibitem [{\citenamefont {Morelli}\ \emph {et~al.}(2020)\citenamefont {Morelli}, \citenamefont {Kl{\"o}ckl}, \citenamefont {Eltschka}, \citenamefont {Siewert},\ and\ \citenamefont {Huber}}]{Morelli2020}%
  \BibitemOpen
  \bibfield  {author} {\bibinfo {author} {\bibfnamefont {S.}~\bibnamefont {Morelli}}, \bibinfo {author} {\bibfnamefont {C.}~\bibnamefont {Kl{\"o}ckl}}, \bibinfo {author} {\bibfnamefont {C.}~\bibnamefont {Eltschka}}, \bibinfo {author} {\bibfnamefont {J.}~\bibnamefont {Siewert}},\ and\ \bibinfo {author} {\bibfnamefont {M.}~\bibnamefont {Huber}},\ }\bibfield  {title} {\bibinfo {title} {Dimensionally sharp inequalities for the linear entropy},\ }\href@noop {} {\bibfield  {journal} {\bibinfo  {journal} {Linear Algebra and its Applications}\ }\textbf {\bibinfo {volume} {584}},\ \bibinfo {pages} {294} (\bibinfo {year} {2020})}\BibitemShut {NoStop}%
\bibitem [{\citenamefont {Pauletti}\ \emph {et~al.}(2024)\citenamefont {Pauletti}, \citenamefont {Silva}, \citenamefont {Canella},\ and\ \citenamefont {França}}]{Pauletti2024}%
  \BibitemOpen
  \bibfield  {author} {\bibinfo {author} {\bibfnamefont {T.}~\bibnamefont {Pauletti}}, \bibinfo {author} {\bibfnamefont {M.}~\bibnamefont {Silva}}, \bibinfo {author} {\bibfnamefont {G.}~\bibnamefont {Canella}},\ and\ \bibinfo {author} {\bibfnamefont {V.}~\bibnamefont {França}},\ }\bibfield  {title} {\bibinfo {title} {Linear entropy fails to predict entanglement behavior in low-density fermionic systems},\ }\href {https://doi.org/https://doi.org/10.1016/j.physa.2024.129824} {\bibfield  {journal} {\bibinfo  {journal} {Physica A: Statistical Mechanics and its Applications}\ }\textbf {\bibinfo {volume} {644}},\ \bibinfo {pages} {129824} (\bibinfo {year} {2024})}\BibitemShut {NoStop}%
\bibitem [{\citenamefont {Lahini}\ \emph {et~al.}(2010)\citenamefont {Lahini}, \citenamefont {Bromberg}, \citenamefont {Christodoulides},\ and\ \citenamefont {Silberberg}}]{Lahini2010}%
  \BibitemOpen
  \bibfield  {author} {\bibinfo {author} {\bibfnamefont {Y.}~\bibnamefont {Lahini}}, \bibinfo {author} {\bibfnamefont {Y.}~\bibnamefont {Bromberg}}, \bibinfo {author} {\bibfnamefont {D.~N.}\ \bibnamefont {Christodoulides}},\ and\ \bibinfo {author} {\bibfnamefont {Y.}~\bibnamefont {Silberberg}},\ }\bibfield  {title} {\bibinfo {title} {Quantum correlations in two-particle anderson localization},\ }\href {https://doi.org/10.1103/PhysRevLett.105.163905} {\bibfield  {journal} {\bibinfo  {journal} {Phys. Rev. Lett.}\ }\textbf {\bibinfo {volume} {105}},\ \bibinfo {pages} {163905} (\bibinfo {year} {2010})}\BibitemShut {NoStop}%
\bibitem [{\citenamefont {Lahini}\ \emph {et~al.}(2012)\citenamefont {Lahini}, \citenamefont {Verbin}, \citenamefont {Huber}, \citenamefont {Bromberg}, \citenamefont {Pugatch},\ and\ \citenamefont {Silberberg}}]{Lahini2012}%
  \BibitemOpen
  \bibfield  {author} {\bibinfo {author} {\bibfnamefont {Y.}~\bibnamefont {Lahini}}, \bibinfo {author} {\bibfnamefont {M.}~\bibnamefont {Verbin}}, \bibinfo {author} {\bibfnamefont {S.~D.}\ \bibnamefont {Huber}}, \bibinfo {author} {\bibfnamefont {Y.}~\bibnamefont {Bromberg}}, \bibinfo {author} {\bibfnamefont {R.}~\bibnamefont {Pugatch}},\ and\ \bibinfo {author} {\bibfnamefont {Y.}~\bibnamefont {Silberberg}},\ }\bibfield  {title} {\bibinfo {title} {Quantum walk of two interacting bosons},\ }\href {https://doi.org/10.1103/PhysRevA.86.011603} {\bibfield  {journal} {\bibinfo  {journal} {Phys. Rev. A}\ }\textbf {\bibinfo {volume} {86}},\ \bibinfo {pages} {011603} (\bibinfo {year} {2012})}\BibitemShut {NoStop}%
\bibitem [{\citenamefont {Oliveira}\ \emph {et~al.}(2023)\citenamefont {Oliveira}, \citenamefont {de~Moura}, \citenamefont {Souza}, \citenamefont {Lyra},\ and\ \citenamefont {Almeida}}]{oliveira2023}%
  \BibitemOpen
  \bibfield  {author} {\bibinfo {author} {\bibfnamefont {M.~F.~V.}\ \bibnamefont {Oliveira}}, \bibinfo {author} {\bibfnamefont {F.~A. B.~F.}\ \bibnamefont {de~Moura}}, \bibinfo {author} {\bibfnamefont {A.~M.~C.}\ \bibnamefont {Souza}}, \bibinfo {author} {\bibfnamefont {M.~L.}\ \bibnamefont {Lyra}},\ and\ \bibinfo {author} {\bibfnamefont {G.~M.~A.}\ \bibnamefont {Almeida}},\ }\bibfield  {title} {\bibinfo {title} {Non-rayleigh signal of interacting quantum particles},\ }\href {https://doi.org/10.1103/PhysRevA.108.023520} {\bibfield  {journal} {\bibinfo  {journal} {Phys. Rev. A}\ }\textbf {\bibinfo {volume} {108}},\ \bibinfo {pages} {023520} (\bibinfo {year} {2023})}\BibitemShut {NoStop}%
\bibitem [{\citenamefont {Shepelyansky}(1994)}]{Shepelyansky1994}%
  \BibitemOpen
  \bibfield  {author} {\bibinfo {author} {\bibfnamefont {D.~L.}\ \bibnamefont {Shepelyansky}},\ }\bibfield  {title} {\bibinfo {title} {Coherent propagation of two interacting particles in a random potential},\ }\href {https://doi.org/10.1103/PhysRevLett.73.2607} {\bibfield  {journal} {\bibinfo  {journal} {Phys. Rev. Lett.}\ }\textbf {\bibinfo {volume} {73}},\ \bibinfo {pages} {2607} (\bibinfo {year} {1994})}\BibitemShut {NoStop}%
\bibitem [{\citenamefont {Bromberg}\ \emph {et~al.}(2009)\citenamefont {Bromberg}, \citenamefont {Lahini}, \citenamefont {Morandotti},\ and\ \citenamefont {Silberberg}}]{Bromberg2009}%
  \BibitemOpen
  \bibfield  {author} {\bibinfo {author} {\bibfnamefont {Y.}~\bibnamefont {Bromberg}}, \bibinfo {author} {\bibfnamefont {Y.}~\bibnamefont {Lahini}}, \bibinfo {author} {\bibfnamefont {R.}~\bibnamefont {Morandotti}},\ and\ \bibinfo {author} {\bibfnamefont {Y.}~\bibnamefont {Silberberg}},\ }\bibfield  {title} {\bibinfo {title} {Quantum and classical correlations in waveguide lattices},\ }\href {https://doi.org/10.1103/PhysRevLett.102.253904} {\bibfield  {journal} {\bibinfo  {journal} {Phys. Rev. Lett.}\ }\textbf {\bibinfo {volume} {102}},\ \bibinfo {pages} {253904} (\bibinfo {year} {2009})}\BibitemShut {NoStop}%
\bibitem [{\citenamefont {Lee}\ \emph {et~al.}(2014)\citenamefont {Lee}, \citenamefont {Rai}, \citenamefont {Noh},\ and\ \citenamefont {Angelakis}}]{Lee2014}%
  \BibitemOpen
  \bibfield  {author} {\bibinfo {author} {\bibfnamefont {C.}~\bibnamefont {Lee}}, \bibinfo {author} {\bibfnamefont {A.}~\bibnamefont {Rai}}, \bibinfo {author} {\bibfnamefont {C.}~\bibnamefont {Noh}},\ and\ \bibinfo {author} {\bibfnamefont {D.~G.}\ \bibnamefont {Angelakis}},\ }\bibfield  {title} {\bibinfo {title} {Probing the effect of interaction in anderson localization using linear photonic lattices},\ }\href {https://doi.org/10.1103/PhysRevA.89.023823} {\bibfield  {journal} {\bibinfo  {journal} {Phys. Rev. A}\ }\textbf {\bibinfo {volume} {89}},\ \bibinfo {pages} {023823} (\bibinfo {year} {2014})}\BibitemShut {NoStop}%
\bibitem [{\citenamefont {Ferreira}\ \emph {et~al.}(2022)\citenamefont {Ferreira}, \citenamefont {Maciel}, \citenamefont {Vianna},\ and\ \citenamefont {Iemini}}]{Ferreira2022}%
  \BibitemOpen
  \bibfield  {author} {\bibinfo {author} {\bibfnamefont {D.~L.~B.}\ \bibnamefont {Ferreira}}, \bibinfo {author} {\bibfnamefont {T.~O.}\ \bibnamefont {Maciel}}, \bibinfo {author} {\bibfnamefont {R.~O.}\ \bibnamefont {Vianna}},\ and\ \bibinfo {author} {\bibfnamefont {F.}~\bibnamefont {Iemini}},\ }\bibfield  {title} {\bibinfo {title} {Quantum correlations, entanglement spectrum, and coherence of the two-particle reduced density matrix in the extended hubbard model},\ }\href {https://doi.org/10.1103/PhysRevB.105.115145} {\bibfield  {journal} {\bibinfo  {journal} {Phys. Rev. B}\ }\textbf {\bibinfo {volume} {105}},\ \bibinfo {pages} {115145} (\bibinfo {year} {2022})}\BibitemShut {NoStop}%
\bibitem [{\citenamefont {Peruzzo}\ \emph {et~al.}(2010)\citenamefont {Peruzzo}, \citenamefont {Lobino}, \citenamefont {Matthews}, \citenamefont {Matsuda}, \citenamefont {Politi}, \citenamefont {Poulios}, \citenamefont {Zhou}, \citenamefont {Lahini}, \citenamefont {Ismail}, \citenamefont {Wörhoff}, \citenamefont {Bromberg}, \citenamefont {Silberberg}, \citenamefont {Thompson},\ and\ \citenamefont {OBrien}}]{Peruzzo2010}%
  \BibitemOpen
  \bibfield  {author} {\bibinfo {author} {\bibfnamefont {A.}~\bibnamefont {Peruzzo}}, \bibinfo {author} {\bibfnamefont {M.}~\bibnamefont {Lobino}}, \bibinfo {author} {\bibfnamefont {J.~C.~F.}\ \bibnamefont {Matthews}}, \bibinfo {author} {\bibfnamefont {N.}~\bibnamefont {Matsuda}}, \bibinfo {author} {\bibfnamefont {A.}~\bibnamefont {Politi}}, \bibinfo {author} {\bibfnamefont {K.}~\bibnamefont {Poulios}}, \bibinfo {author} {\bibfnamefont {X.-Q.}\ \bibnamefont {Zhou}}, \bibinfo {author} {\bibfnamefont {Y.}~\bibnamefont {Lahini}}, \bibinfo {author} {\bibfnamefont {N.}~\bibnamefont {Ismail}}, \bibinfo {author} {\bibfnamefont {K.}~\bibnamefont {Wörhoff}}, \bibinfo {author} {\bibfnamefont {Y.}~\bibnamefont {Bromberg}}, \bibinfo {author} {\bibfnamefont {Y.}~\bibnamefont {Silberberg}}, \bibinfo {author} {\bibfnamefont {M.~G.}\ \bibnamefont {Thompson}},\ and\ \bibinfo {author} {\bibfnamefont {J.~L.}\ \bibnamefont {OBrien}},\ }\bibfield  {title} {\bibinfo {title} {Quantum walks of correlated photons},\ }\href
  {https://doi.org/10.1126/science.1193515} {\bibfield  {journal} {\bibinfo  {journal} {Science}\ }\textbf {\bibinfo {volume} {329}},\ \bibinfo {pages} {1500} (\bibinfo {year} {2010})}\BibitemShut {NoStop}%
\bibitem [{\citenamefont {Krimer}\ and\ \citenamefont {Khomeriki}(2011)}]{Krimer2011}%
  \BibitemOpen
  \bibfield  {author} {\bibinfo {author} {\bibfnamefont {D.~O.}\ \bibnamefont {Krimer}}\ and\ \bibinfo {author} {\bibfnamefont {R.}~\bibnamefont {Khomeriki}},\ }\bibfield  {title} {\bibinfo {title} {Realization of discrete quantum billiards in a two-dimensional optical lattice},\ }\href {https://doi.org/10.1103/PhysRevA.84.041807} {\bibfield  {journal} {\bibinfo  {journal} {Phys. Rev. A}\ }\textbf {\bibinfo {volume} {84}},\ \bibinfo {pages} {041807} (\bibinfo {year} {2011})}\BibitemShut {NoStop}%
\bibitem [{\citenamefont {Longhi}\ and\ \citenamefont {Valle}(2011)}]{Longhi2011}%
  \BibitemOpen
  \bibfield  {author} {\bibinfo {author} {\bibfnamefont {S.}~\bibnamefont {Longhi}}\ and\ \bibinfo {author} {\bibfnamefont {G.~D.}\ \bibnamefont {Valle}},\ }\bibfield  {title} {\bibinfo {title} {Tunneling control of strongly correlated particles on a lattice: a photonic realization},\ }\href {https://doi.org/10.1364/OL.36.004743} {\bibfield  {journal} {\bibinfo  {journal} {Opt. Lett.}\ }\textbf {\bibinfo {volume} {36}},\ \bibinfo {pages} {4743} (\bibinfo {year} {2011})}\BibitemShut {NoStop}%
\bibitem [{\citenamefont {Corrielli}\ \emph {et~al.}(2013)\citenamefont {Corrielli}, \citenamefont {Crespi}, \citenamefont {Della~Valle}, \citenamefont {Longhi},\ and\ \citenamefont {Osellame}}]{Corrielli2013}%
  \BibitemOpen
  \bibfield  {author} {\bibinfo {author} {\bibfnamefont {G.}~\bibnamefont {Corrielli}}, \bibinfo {author} {\bibfnamefont {A.}~\bibnamefont {Crespi}}, \bibinfo {author} {\bibfnamefont {G.}~\bibnamefont {Della~Valle}}, \bibinfo {author} {\bibfnamefont {S.}~\bibnamefont {Longhi}},\ and\ \bibinfo {author} {\bibfnamefont {R.}~\bibnamefont {Osellame}},\ }\bibfield  {title} {\bibinfo {title} {Fractional bloch oscillations in photonic lattices},\ }\href {https://doi.org/10.1038/ncomms2578} {\bibfield  {journal} {\bibinfo  {journal} {Nature Communications}\ }\textbf {\bibinfo {volume} {4}},\ \bibinfo {pages} {1555} (\bibinfo {year} {2013})}\BibitemShut {NoStop}%
\bibitem [{\citenamefont {Baumgratz}\ \emph {et~al.}(2014)\citenamefont {Baumgratz}, \citenamefont {Cramer},\ and\ \citenamefont {Plenio}}]{Baumgratz2014}%
  \BibitemOpen
  \bibfield  {author} {\bibinfo {author} {\bibfnamefont {T.}~\bibnamefont {Baumgratz}}, \bibinfo {author} {\bibfnamefont {M.}~\bibnamefont {Cramer}},\ and\ \bibinfo {author} {\bibfnamefont {M.~B.}\ \bibnamefont {Plenio}},\ }\bibfield  {title} {\bibinfo {title} {Quantifying coherence},\ }\href {https://doi.org/10.1103/PhysRevLett.113.140401} {\bibfield  {journal} {\bibinfo  {journal} {Phys. Rev. Lett.}\ }\textbf {\bibinfo {volume} {113}},\ \bibinfo {pages} {140401} (\bibinfo {year} {2014})}\BibitemShut {NoStop}%
\bibitem [{\citenamefont {Dias}\ \emph {et~al.}(2007)\citenamefont {Dias}, \citenamefont {Nascimento}, \citenamefont {Lyra},\ and\ \citenamefont {de~Moura}}]{Dias2007}%
  \BibitemOpen
  \bibfield  {author} {\bibinfo {author} {\bibfnamefont {W.~S.}\ \bibnamefont {Dias}}, \bibinfo {author} {\bibfnamefont {E.~M.}\ \bibnamefont {Nascimento}}, \bibinfo {author} {\bibfnamefont {M.~L.}\ \bibnamefont {Lyra}},\ and\ \bibinfo {author} {\bibfnamefont {F.~A. B.~F.}\ \bibnamefont {de~Moura}},\ }\bibfield  {title} {\bibinfo {title} {Frequency doubling of bloch oscillations for interacting electrons in a static electric field},\ }\href {https://doi.org/10.1103/PhysRevB.76.155124} {\bibfield  {journal} {\bibinfo  {journal} {Phys. Rev. B}\ }\textbf {\bibinfo {volume} {76}},\ \bibinfo {pages} {155124} (\bibinfo {year} {2007})}\BibitemShut {NoStop}%
\bibitem [{\citenamefont {Dias}\ \emph {et~al.}(2010)\citenamefont {Dias}, \citenamefont {Lyra},\ and\ \citenamefont {{de Moura}}}]{Dias2010}%
  \BibitemOpen
  \bibfield  {author} {\bibinfo {author} {\bibfnamefont {W.}~\bibnamefont {Dias}}, \bibinfo {author} {\bibfnamefont {M.}~\bibnamefont {Lyra}},\ and\ \bibinfo {author} {\bibfnamefont {F.}~\bibnamefont {{de Moura}}},\ }\bibfield  {title} {\bibinfo {title} {The role of hubbard-like interaction in the dynamics of two interacting electrons},\ }\href {https://doi.org/https://doi.org/10.1016/j.physleta.2010.09.005} {\bibfield  {journal} {\bibinfo  {journal} {Physics Letters A}\ }\textbf {\bibinfo {volume} {374}},\ \bibinfo {pages} {4554} (\bibinfo {year} {2010})}\BibitemShut {NoStop}%
\bibitem [{\citenamefont {Buchleitner}\ and\ \citenamefont {Kolovsky}(2003)}]{Buchleitner2003}%
  \BibitemOpen
  \bibfield  {author} {\bibinfo {author} {\bibfnamefont {A.}~\bibnamefont {Buchleitner}}\ and\ \bibinfo {author} {\bibfnamefont {A.~R.}\ \bibnamefont {Kolovsky}},\ }\bibfield  {title} {\bibinfo {title} {Interaction-induced decoherence of atomic bloch oscillations},\ }\href {https://doi.org/10.1103/PhysRevLett.91.253002} {\bibfield  {journal} {\bibinfo  {journal} {Phys. Rev. Lett.}\ }\textbf {\bibinfo {volume} {91}},\ \bibinfo {pages} {253002} (\bibinfo {year} {2003})}\BibitemShut {NoStop}%
\bibitem [{\citenamefont {Rapedius}\ and\ \citenamefont {Korsch}(2012)}]{Rapedius2012}%
  \BibitemOpen
  \bibfield  {author} {\bibinfo {author} {\bibfnamefont {K.}~\bibnamefont {Rapedius}}\ and\ \bibinfo {author} {\bibfnamefont {H.~J.}\ \bibnamefont {Korsch}},\ }\bibfield  {title} {\bibinfo {title} {Interaction-induced decoherence in non-hermitian quantum walks of ultracold bosons},\ }\href {https://doi.org/10.1103/PhysRevA.86.025601} {\bibfield  {journal} {\bibinfo  {journal} {Phys. Rev. A}\ }\textbf {\bibinfo {volume} {86}},\ \bibinfo {pages} {025601} (\bibinfo {year} {2012})}\BibitemShut {NoStop}%
\bibitem [{\citenamefont {Benatti}\ \emph {et~al.}(2020)\citenamefont {Benatti}, \citenamefont {Floreanini}, \citenamefont {Franchini},\ and\ \citenamefont {Marzolino}}]{Benatti2020}%
  \BibitemOpen
  \bibfield  {author} {\bibinfo {author} {\bibfnamefont {F.}~\bibnamefont {Benatti}}, \bibinfo {author} {\bibfnamefont {R.}~\bibnamefont {Floreanini}}, \bibinfo {author} {\bibfnamefont {F.}~\bibnamefont {Franchini}},\ and\ \bibinfo {author} {\bibfnamefont {U.}~\bibnamefont {Marzolino}},\ }\bibfield  {title} {\bibinfo {title} {Entanglement in indistinguishable particle systems},\ }\href {https://doi.org/https://doi.org/10.1016/j.physrep.2020.07.003} {\bibfield  {journal} {\bibinfo  {journal} {Physics Reports}\ }\textbf {\bibinfo {volume} {878}},\ \bibinfo {pages} {1} (\bibinfo {year} {2020})},\ \bibinfo {note} {entanglement in indistinguishable particle systems}\BibitemShut {NoStop}%
\bibitem [{\citenamefont {Calabrese}\ and\ \citenamefont {Cardy}(2006)}]{Calabrese2006}%
  \BibitemOpen
  \bibfield  {author} {\bibinfo {author} {\bibfnamefont {P.}~\bibnamefont {Calabrese}}\ and\ \bibinfo {author} {\bibfnamefont {J.}~\bibnamefont {Cardy}},\ }\bibfield  {title} {\bibinfo {title} {Time dependence of correlation functions following a quantum quench},\ }\href {https://doi.org/10.1103/PhysRevLett.96.136801} {\bibfield  {journal} {\bibinfo  {journal} {Phys. Rev. Lett.}\ }\textbf {\bibinfo {volume} {96}},\ \bibinfo {pages} {136801} (\bibinfo {year} {2006})}\BibitemShut {NoStop}%
\bibitem [{\citenamefont {Streltsov}\ \emph {et~al.}(2015)\citenamefont {Streltsov}, \citenamefont {Singh}, \citenamefont {Dhar}, \citenamefont {Bera},\ and\ \citenamefont {Adesso}}]{Streltsov2015}%
  \BibitemOpen
  \bibfield  {author} {\bibinfo {author} {\bibfnamefont {A.}~\bibnamefont {Streltsov}}, \bibinfo {author} {\bibfnamefont {U.}~\bibnamefont {Singh}}, \bibinfo {author} {\bibfnamefont {H.~S.}\ \bibnamefont {Dhar}}, \bibinfo {author} {\bibfnamefont {M.~N.}\ \bibnamefont {Bera}},\ and\ \bibinfo {author} {\bibfnamefont {G.}~\bibnamefont {Adesso}},\ }\bibfield  {title} {\bibinfo {title} {Measuring quantum coherence with entanglement},\ }\href {https://doi.org/10.1103/PhysRevLett.115.020403} {\bibfield  {journal} {\bibinfo  {journal} {Phys. Rev. Lett.}\ }\textbf {\bibinfo {volume} {115}},\ \bibinfo {pages} {020403} (\bibinfo {year} {2015})}\BibitemShut {NoStop}%
\bibitem [{\citenamefont {Salath\'e}\ \emph {et~al.}(2015)\citenamefont {Salath\'e}, \citenamefont {Mondal}, \citenamefont {Oppliger}, \citenamefont {Heinsoo}, \citenamefont {Kurpiers}, \citenamefont {Poto\ifmmode~\check{c}\else \v{c}\fi{}nik}, \citenamefont {Mezzacapo}, \citenamefont {Las~Heras}, \citenamefont {Lamata}, \citenamefont {Solano}, \citenamefont {Filipp},\ and\ \citenamefont {Wallraff}}]{Salathe2015}%
  \BibitemOpen
  \bibfield  {author} {\bibinfo {author} {\bibfnamefont {Y.}~\bibnamefont {Salath\'e}}, \bibinfo {author} {\bibfnamefont {M.}~\bibnamefont {Mondal}}, \bibinfo {author} {\bibfnamefont {M.}~\bibnamefont {Oppliger}}, \bibinfo {author} {\bibfnamefont {J.}~\bibnamefont {Heinsoo}}, \bibinfo {author} {\bibfnamefont {P.}~\bibnamefont {Kurpiers}}, \bibinfo {author} {\bibfnamefont {A.}~\bibnamefont {Poto\ifmmode~\check{c}\else \v{c}\fi{}nik}}, \bibinfo {author} {\bibfnamefont {A.}~\bibnamefont {Mezzacapo}}, \bibinfo {author} {\bibfnamefont {U.}~\bibnamefont {Las~Heras}}, \bibinfo {author} {\bibfnamefont {L.}~\bibnamefont {Lamata}}, \bibinfo {author} {\bibfnamefont {E.}~\bibnamefont {Solano}}, \bibinfo {author} {\bibfnamefont {S.}~\bibnamefont {Filipp}},\ and\ \bibinfo {author} {\bibfnamefont {A.}~\bibnamefont {Wallraff}},\ }\bibfield  {title} {\bibinfo {title} {Digital quantum simulation of spin models with circuit quantum electrodynamics},\ }\href {https://doi.org/10.1103/PhysRevX.5.021027} {\bibfield  {journal}
  {\bibinfo  {journal} {Phys. Rev. X}\ }\textbf {\bibinfo {volume} {5}},\ \bibinfo {pages} {021027} (\bibinfo {year} {2015})}\BibitemShut {NoStop}%
\bibitem [{\citenamefont {Yan}\ \emph {et~al.}(2019)\citenamefont {Yan}, \citenamefont {Zhang}, \citenamefont {Gong}, \citenamefont {Wu}, \citenamefont {Zheng}, \citenamefont {Li}, \citenamefont {Wang}, \citenamefont {Liang}, \citenamefont {Lin}, \citenamefont {Xu}, \citenamefont {Guo}, \citenamefont {Sun}, \citenamefont {Peng}, \citenamefont {Xia}, \citenamefont {Deng}, \citenamefont {Rong}, \citenamefont {You}, \citenamefont {Nori}, \citenamefont {Fan}, \citenamefont {Zhu},\ and\ \citenamefont {Pan}}]{yan19}%
  \BibitemOpen
  \bibfield  {author} {\bibinfo {author} {\bibfnamefont {Z.}~\bibnamefont {Yan}}, \bibinfo {author} {\bibfnamefont {Y.-R.}\ \bibnamefont {Zhang}}, \bibinfo {author} {\bibfnamefont {M.}~\bibnamefont {Gong}}, \bibinfo {author} {\bibfnamefont {Y.}~\bibnamefont {Wu}}, \bibinfo {author} {\bibfnamefont {Y.}~\bibnamefont {Zheng}}, \bibinfo {author} {\bibfnamefont {S.}~\bibnamefont {Li}}, \bibinfo {author} {\bibfnamefont {C.}~\bibnamefont {Wang}}, \bibinfo {author} {\bibfnamefont {F.}~\bibnamefont {Liang}}, \bibinfo {author} {\bibfnamefont {J.}~\bibnamefont {Lin}}, \bibinfo {author} {\bibfnamefont {Y.}~\bibnamefont {Xu}}, \bibinfo {author} {\bibfnamefont {C.}~\bibnamefont {Guo}}, \bibinfo {author} {\bibfnamefont {L.}~\bibnamefont {Sun}}, \bibinfo {author} {\bibfnamefont {C.-Z.}\ \bibnamefont {Peng}}, \bibinfo {author} {\bibfnamefont {K.}~\bibnamefont {Xia}}, \bibinfo {author} {\bibfnamefont {H.}~\bibnamefont {Deng}}, \bibinfo {author} {\bibfnamefont {H.}~\bibnamefont {Rong}}, \bibinfo {author} {\bibfnamefont {J.~Q.}\
  \bibnamefont {You}}, \bibinfo {author} {\bibfnamefont {F.}~\bibnamefont {Nori}}, \bibinfo {author} {\bibfnamefont {H.}~\bibnamefont {Fan}}, \bibinfo {author} {\bibfnamefont {X.}~\bibnamefont {Zhu}},\ and\ \bibinfo {author} {\bibfnamefont {J.-W.}\ \bibnamefont {Pan}},\ }\bibfield  {title} {\bibinfo {title} {Strongly correlated quantum walks with a 12-qubit superconducting processor},\ }\href {https://doi.org/10.1126/science.aaw1611} {\bibfield  {journal} {\bibinfo  {journal} {Science}\ }\textbf {\bibinfo {volume} {364}},\ \bibinfo {pages} {753} (\bibinfo {year} {2019})}\BibitemShut {NoStop}%
\bibitem [{\citenamefont {Bloch}(2005)}]{Bloch2005}%
  \BibitemOpen
  \bibfield  {author} {\bibinfo {author} {\bibfnamefont {I.}~\bibnamefont {Bloch}},\ }\bibfield  {title} {\bibinfo {title} {Ultracold quantum gases in optical lattices},\ }\href {https://doi.org/10.1038/nphys138} {\bibfield  {journal} {\bibinfo  {journal} {Nature Physics}\ }\textbf {\bibinfo {volume} {1}},\ \bibinfo {pages} {23} (\bibinfo {year} {2005})}\BibitemShut {NoStop}%
\bibitem [{\citenamefont {Gross}\ and\ \citenamefont {Bloch}(2017)}]{Gross2017}%
  \BibitemOpen
  \bibfield  {author} {\bibinfo {author} {\bibfnamefont {C.}~\bibnamefont {Gross}}\ and\ \bibinfo {author} {\bibfnamefont {I.}~\bibnamefont {Bloch}},\ }\bibfield  {title} {\bibinfo {title} {Quantum simulations with ultracold atoms in optical lattices},\ }\href {https://doi.org/10.1126/science.aal3837} {\bibfield  {journal} {\bibinfo  {journal} {Science}\ }\textbf {\bibinfo {volume} {357}},\ \bibinfo {pages} {995} (\bibinfo {year} {2017})}\BibitemShut {NoStop}%
\bibitem [{\citenamefont {Quan}\ and\ \citenamefont {Zurek}(2010)}]{Quan2010}%
  \BibitemOpen
  \bibfield  {author} {\bibinfo {author} {\bibfnamefont {H.~T.}\ \bibnamefont {Quan}}\ and\ \bibinfo {author} {\bibfnamefont {W.~H.}\ \bibnamefont {Zurek}},\ }\bibfield  {title} {\bibinfo {title} {Testing quantum adiabaticity with quench echo},\ }\href {https://doi.org/10.1088/1367-2630/12/9/093025} {\bibfield  {journal} {\bibinfo  {journal} {New Journal of Physics}\ }\textbf {\bibinfo {volume} {12}},\ \bibinfo {pages} {093025} (\bibinfo {year} {2010})}\BibitemShut {NoStop}%
\end{thebibliography}

%

\end{document}